\documentclass[apsf,pra,showpacs,amsmath,amssymb,twocolumn,nofootinbib,superscriptaddress,longbibliography]{revtex4-2}

\usepackage[pdftex]{graphicx}

\usepackage{mathrsfs,amsmath,amssymb,amsthm,graphicx}
\usepackage{fancyhdr}
\usepackage{caption}
\usepackage{subcaption}

\usepackage{braket}
\usepackage[colorlinks=true,
            linkcolor=red,
            urlcolor=blue,
            citecolor=blue]{hyperref}
 
\draft 

\begin{document}
\bibliographystyle{apsrev}

\title{Quantum Time Transfer: A Practical Method for Lossy and Noisy Channels}
\author{Randy Lafler}
\affiliation{Air Force Research Laboratory, Directed Energy Directorate, Kirtland AFB, NM, United States}

\email[AFRL.RDSS.OrgMailbox@us.af.mil\\
DISTRIBUTION A: Approved for public release; distribution is unlimited. \\
Public Affairs release approval AFRL-2023-2447]{}

\author{R. Nicholas Lanning}
\affiliation{Air Force Research Laboratory, Directed Energy Directorate, Kirtland AFB, NM, United States}

\date{\today}

\begin{abstract}
Timing requirements for long-range quantum networking are driven by the necessity of synchronizing the arrival of photons, from independent sources, for Bell-state measurements.
Thus, characteristics such as repetition rate and pulse duration influence the precision required to enable quantum networking tasks such as teleportation and entanglement swapping. 
Some solutions have been proposed utilizing classical laser pulses, frequency combs, and biphoton sources.
In this article, we explore the utility of the latter method since it is based upon quantum  phenomena, which makes it naturally covert, and potentially quantum secure. 
Furthermore, it can utilize relatively low performance quantum-photon sources and detection equipment, but provides picosecond-level timing precision even under high loss and high noise channel conditions representative of daytime space-Earth links.
Therefore, this method is relevant for daytime space-Earth quantum networking and/or providing high-precision secure timing in GPS denied environments.
\end{abstract}

\maketitle 
\pagestyle{fancy}
\cfoot{DISTRIBUTION A: Approved for public release; distribution is unlimited. Public Affairs release approval AFRL-2023-2447.}
\lhead{}
\chead{}
\rhead{\thepage}

\section{Introduction} \label{sec: Introduction}
Precise synchronization of remote clocks is at the heart of position, navigation, and timing (PNT), high-speed transactions, distributed computing, and as of late, quantum networking.
To enable \textit{global-scale} quantum networking, one needs to distribute entanglement between distant quantum nodes \cite{wehner2018quantum, van2014quantum}, which will likely require a series of entanglement swapping operations between different arrangements of ground and satellite quantum nodes \cite{aspelmeyer2003long,boone2015entanglement}.
If ultra-narrow spectral filtering is utilized, then each entanglement swapping operation could require Bell-state measurements with as little as nanosecond-scale timing precision.
This can be achieved utilizing single-photon avalanche-diode (SPAD) detectors and synchronization provided by current global positioning system (GPS) public signals \cite{lombardi2001time}.
However, ultra-narrow spectral filtering is not currently compatible with entangled photon sources, which are generally broader band, and would result in considerable attenuation. 
An alternate approach is to increase the timing precision to the picosecond or femtosecond level, along with choosing the appropriate pulse duration and spectral-temporal filtering.
This would result in a higher probability of success per pulse and maintain the rejection of noise photons that scatter into the channel.
Therefore, techniques to precisely synchronize remote clocks are an important ongoing area of research.

Perhaps the most straight forward optical-time-transfer technique uses laser pulses, photodetectors, and software-based correlation methods.
For example, time transfer by laser link (T2L2) demonstrations have achieved picosecond-scale precision between remote ground stations operating in common view with the Jason-2 satellite \cite{exertier2014time}.
In contrast, there are also hardware-based methods such as optical two-way time and frequency transfer (O-TWTFT) \cite{giorgetta2013optical,sinclair2016synchronization}, which utilizes frequency combs and linear optical sampling to synchronize two remote clocks to femtosecond precision.  
Demonstrations of O-TWTFT have been performed between stationary sites \cite{giorgetta2013optical,sinclair2016synchronization} or slow moving drones, $<25$ m/s.
The conclusions in Refs. \cite{sinclair2019femtosecond,swann2019measurement} suggest that O-TWTFT can maintain femtosecond-scale precision despite the high orbital velocities and non-reciprocity of two-way Earth-satellite links, but a demonstration over channel conditions representative of an Earth-satellite link has yet to be performed.

The concept of using quantum phenomena has also emerged as a possible solution for precise synchronization.
One example is the Earth-satellite synchronization demonstration that used quantum key distribution (QKD) with attenuated laser pulses and a high powered sync pulse \cite{dai2020towards}.  
However, this technique relies on extraneous components and classical pulsing making it more complex and less covert.
Another example relies on Hong-Oh-Mandel (HOM) quantum interference between entangled photon pairs \cite{quan2016demonstration}.
However, it is challenging to utilize this technique over freespace-atmospheric channels due to aberrations of the photon transverse momentum and the very high level of attenuation characteristic of the double-pass geometry.

Another technique consists of utilizing the femtosecond-scale temporal correlations of photon pairs created in spontaneous-parametric-down conversion (SPDC) photon sources \cite{valencia2004distant}. 
In this case, the relative time offset between two remote clocks is measured with the following procedure: 
(1)~a series of photon pairs are separated and transmitted to two remote sites,
(2)~the photons are detected and time tagged based on the respective local clock, 
(3)~after enough detection events are collected, the series of arrival times from each site are combined and correlation methods are used to find the clock offset.  
This technique was demonstrated in Refs. \cite{valencia2004distant,ho2009clock,lee2019symmetrical} and it is referred to as quantum time transfer (QTT) for the remainder of this paper.
One-way QTT \cite{ho2009clock} can provide relative clock synchronization and two-way QTT \cite{lee2019symmetrical} can provide absolute clock synchronization.

One can think of QTT as a quantum analog of T2L2, where the laser pulses of T2L2 are replaced with randomly arriving, but correlated, photon pairs.
A continuous wave (CW) biphoton-source creates pairs that are distributed randomly over the acquisition time. 
Therefore, correlating the signals with QTT reveals a single peak corresponding to the clock offset, whereas deterministic pulsing results in a collection of peaks repeating at the pulse period.
It is possible to perform QTT using pulsed-biphoton sources, provided the photon pairs are still created \textit{probabilistically}.
The key requirement is that some of the pulses must be empty and Alice must register that information.
For pulsed-biphoton sources that are probabilistic, one typically keeps the probability of pair creation per pulse low in order to suppress multi-pair events.
This results in many empty pulses, and the coincidence peak corresponding to the clock offset will be larger than the peaks corresponding to the pump-pulse cycle.
Similarly, the algorithm will work for weak-coherent-pulse quantum-key-distribution photon sources, provided that enough random vacuum-decoy pulses are used to suppress the side peaks.

In this article, we propose a simple and computationally fast method of QTT and investigate the CW biphoton-source approach.
In Sec.~\ref{sec: QTT Theory} we introduce a potential freespace architecture, outline the algorithm, and discuss the expected precision.
In Sec.~\ref{sec: Sim} we assume several different heralding efficiency sources and perform a simulation spanning a large space of channel attenuation and noise-photon rates.
We consequently discuss the probability of success per acquisition, the standard error of the mean of the measured clock offset, and the overlapping Allan deviation, which conveys the stability and noise profile of the two clock system.
In Secs.~\ref{sec:FiberLength} and \ref{sec:HEMIS} we interpret the results in the context of fiber channels and space-to-Earth downlinks with sky-noise photons and slant-path turbulence, thereby demonstrating the relevance of this method for global-scale quantum networking.

\section{Quantum Time Transfer}\label{sec: QTT Theory}

\begin{figure}[t!]
	\includegraphics[width=1.0\linewidth]{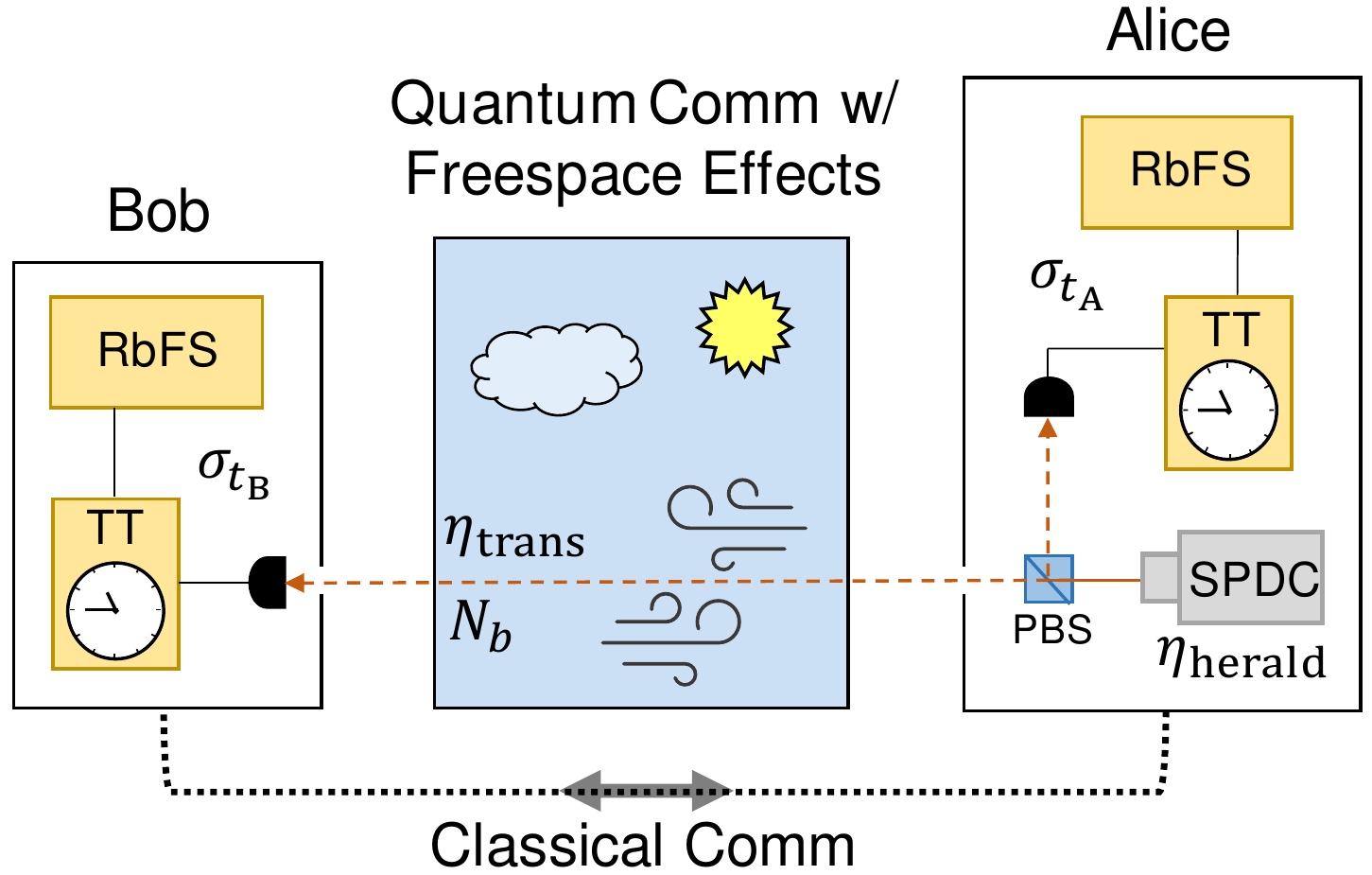}	
	\caption{Diagram of relevant components and phenomena present when performing QTT over daytime atmospheric channels. Alice uses a Type-II SPDC biphoton source to create temporal correlations that facilitate QTT.  She keeps one of the biphotons locally and records its detection time with a detection system consisting of a detector, a time tagger (TT), and Rubidium frequency standard (RbFS), collectively with jitter $\sigma_{t_A}$. With heralding efficiency $\eta_{\mathrm{herald}}$ she sends the companion photon to Bob who similarly records the detection event, subsequent to his total channel attenuation, which includes the atmospheric transmission $\eta_{\mathrm{trans}}$ (see Eq.~\ref{eq:EtaBob}). The atmosphere also causes sky noise photons $N_b$ to scatter into the channel. Bob announces the time tags over a classical channel and Alice uses this information to perform a correlation measurement to determine the relative clock offset $\tau$.}
	\label{fig:schematic}
\end{figure} 
\subsection{Example Architecture}
\begin{figure}[t!]
	\includegraphics[width=0.85\linewidth]{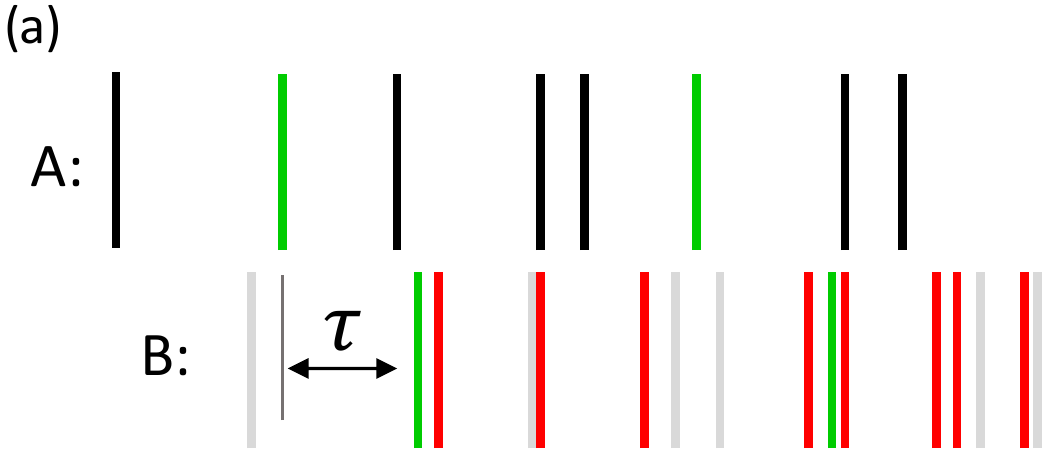}\\
	\vspace{0.0001cm}
	\includegraphics[width=0.85\linewidth]{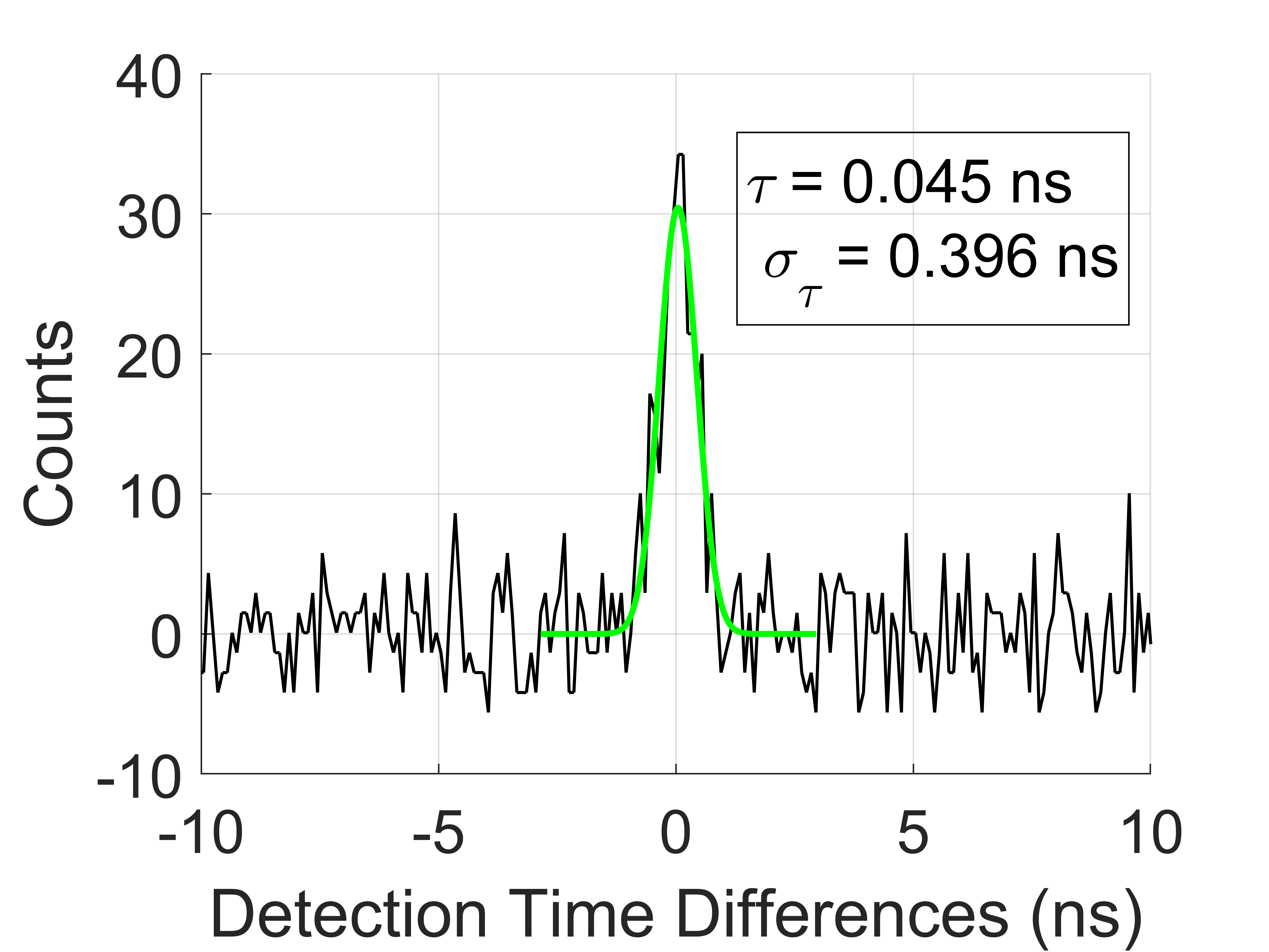}\\
	\caption{(a) Data stream depicting the phenomena underpinning QTT. Alice records a stream of detection events represented by the time tags (see top row). Bob records the same events as Alice with efficiency $\eta_{B}$ (see bottom row). In Bob's stream, potential detection events that are lost are grayed out, and events corresponding to background noise photons are colored red. The infrequent true coincidences that give rise to a correlation peak at $\tau$ are colored green. (b) An example correlation histogram (black curve). The green curve is a Gaussian fit of the correlation peak, which establishes the clock offset $\tau$ and its standard deviation $\sigma_{\tau}$ resulting from the total systematic detection time jitter (see Sec. \ref{sec:Precision}).}
	\label{fig:data_stream}
\end{figure}

QTT has several advantages over the aforementioned techniques that make it promising for \textit{freespace} channels.  
Namely, the architecture is relatively simple, low size, weight, and power (SWAP), and the algorithm is robust to loss and noise photons.
Figure~\ref{fig:schematic} depicts some of the relevant components and phenomena comprising QTT over daytime-freespace channels.
We assume for the moment Alice and Bob both utilize SPAD detectors, time-tagging units, and Rubidium frequency standards (RbFS) to provide initial system stability, which is discussed more in the following subsection.
Alice records one of the biphotons locally subsequent to spectral, detector, and heralding efficiencies
\begin{equation}
\eta_{A} = \eta_{\mathrm{spec}}\eta_{\mathrm{det}}\eta_{\mathrm{herald}}.
\end{equation}
Bob similarly records detection events with extra attenuation imposed by the optical receiver and the atmospheric transmission     
\begin{equation}
\label{eq:EtaBob}
\eta_{B} = \eta_{\mathrm{ch}} \, \eta_{\mathrm{herald}},
\end{equation}
where $\eta_{\mathrm{ch}} = \eta_{\mathrm{spec}}\eta_{\mathrm{det}}\eta_{\mathrm{rec}}\eta_{\mathrm{trans}}$ is Bob's channel attenuation.
Background-noise photons $N_b$, which can be calculated using radiometric equations \cite{er2005background}, scatter to Bob's detector subsequent to attenuation $\eta_{\mathrm{spec}}\eta_{\mathrm{det}}\eta_{\mathrm{rec}}$.
Figure~\ref{fig:data_stream}(a) shows the effect of these phenomena on the time series of detection events registered by Bob.
Alice records a stream of detection events represented by the time tags in the top row. 
Bob records the same events as Alice with efficiency $\eta_{B}$. 
In Bob's stream, potential detection events that are lost are grayed out, and events corresponding to background noise photons are colored red. 
The infrequent true coincidences that give rise to a correlation signal at the clock offset $\tau$ are colored green.

\subsection{Algorithm}\label{sec: QTT Theory Algo}
The QTT algorithm presented in Ref.~\cite{ho2009clock} measures the total clock offset $\tau$ with a series of increasingly precise cross-correlations.
Their algorithm can be summarized as follows.
First, judiciously choose an acquisition time depending on the rate of Alice and Bob's detection events and divide it into an empty array of $N$ bins of width $w$, one array each for Alice and Bob.
Second, for each array assign a 1 to every bin where a time tag is present and a 0 otherwise.
Third, calculate the discrete cross-correlation of the arrays to find a peak that corresponds to the relative offset between the Alice and Bob clocks.
The algorithm can be repeated again with a narrower bin width to increase the measurement precision, but it is very sensitive to the choice of bin size, acquisition time, etc.
In the regime of high channel attenuation, high background rates, and large, unknown $\tau$, we found this algorithm to be difficult to optimize and extremely computationally intensive.
This is because a very large number of bins, that is, $N>2^{24}$ for a 1 second aquisition time, are required to isolate enough ``true" coincidences from the randomly arriving background photons in order to bring the correlation peak above the noise.
Nevertheless, this algorithm can be used in situations with low background rates, such as QTT in optical fiber.

In order to investigate the extreme conditions of daytime-freespace quantum links, we developed the following coincidence finding algorithm that is based on a simple arrival-time-difference histogram.  
Our algorithm is quick, computationally efficient, and maintains sub-nanosecond timing precision over the parameter space we investigate in Sec.~\ref{sec: Sim}:

\begin{itemize}
\item[1.] Concatenate Alice and Bob time-tag arrays, $t_A$ and $t_B$, respectively, and sort the resulting array $M$ from earliest to latest time.
\item[2.] Find the indexes $k$ of $t_B$ in $M$, where $M_{k}$ returns the array $t_B$ of Bob's time tags.
\item[3.] Calculate the time differences $\tau_{ik}$ between each of Bob's time tags $M_{k}$ and the neighboring Alice time tags $M_{k\pm i}$ preceding and trailing each Bob time tag,
\begin{equation}\label{eq:TimeDiff}
\begin{split}
\tau_{ik}^{(\pm)} &=M_{k \pm i}-M_{k},\\
\end{split}
\end{equation}
where the index $i$ goes from $0$ to an upper limit $n$, which must be large enough to contain the offset.
\item[4.] Histogram the resulting time differences $\tau_{ik}$ with bin size $T_{\mathrm{bin}}$ over a range large enough to include the expected clock offset.
\item[5.] When the algorithm is successful, there is a Gaussian feature in the histogram with mean corresponding to the relative clock offset $\tau$ and standard deviation $\sigma_{\tau}$ corresponding to the system jitter. 
\end{itemize}

QTT can be used as a stand-alone protocol, or to underpin other quantum protocols, such as entanglement-based QKD or entanglement distribution.
In the latter case, one would want to find and isolate the coincidence detection events from the background, and use them for further processing. 
To do this, Bob adjusts his time tags $t_B$ according to 
\begin{equation}
\label{eq: Coin Track}
t_B^\prime = (t_B+\tau) \times (1+\Delta U),
\end{equation}
where $\Delta U$ is the clock drift estimated by subtracting successive clock offset measurements $\tau_i$ and dividing by the acquisition time,
\begin{equation}
\label{eq: freq drift}
\Delta U = \frac{\tau_{i+1}-\tau_{i}}{T_{\mathrm{a}}}.
\end{equation}
The procedure outlined in this section, culminating in Eq.~\ref{eq: Coin Track}, is a relative clock synchronization since the time-of-flight of the photons between Alice and Bob is not known with high precision.
If instead one performs QTT in both directions, that is, two-way QTT, the propagation time can be measured and the clocks can be synchronized absolutely \cite{lee2019symmetrical}.

\subsection{Precision} \label{sec:Precision}
Our QTT algorithm generates a Gaussian correlation feature with mean equal to the clock offset $\tau$ and standard deviation $\sigma_{\tau}$. 
Consequently, the central limit theorem suggests that the uncertainty of our clock offset measurement $\tau$ is 
\begin{equation}
\label{eq: SEM equ}
\sigma_{\overline{\tau}} = \sigma_{\tau} / \sqrt{N_{\mathrm{T}}},
\end{equation}
where for the remainder of the paper we refer to $\sigma_{\overline{\tau}}$ as the standard error of the mean (SEM), $N_{\mathrm{T}} = N_{\mathrm{C}}-N_{\mathrm{AC}}$ is an estimate of the number of true coincidences, $N_{\mathrm{C}}$ is the number of measured coincidences, and $N_{\mathrm{AC}}$ is an estimate of the number of accidental coincidences.
The standard deviation of the correlation $\sigma_{\tau}$ is primarily a measure of the systematic timing error of all the detection components in the system
\begin{equation}
\label{eq:syserror}
\sigma^{(\mathrm{sys})}_t = \sqrt{ \sigma^2_{t_\mathrm{A}} + \sigma^2_{t_\mathrm{B}} },
\end{equation}
where $\sigma_{t_\mathrm{A}}$ and $\sigma_{t_\mathrm{B}}$ are the timing jitters of Alice and Bob's systems, respectively.

In Fig.~\ref{fig:data_stream}(b) we present an example correlation motivated by challenging channel conditions and timing jitter observed in our testbed \cite{gruneisen2020adaptive}.
Namely, we assume a 2-Mcps source, channel attenuation $\eta_{\mathrm{ch}}=-23$ dB, background photons $N_b \approx 9 \times 10^5$, and a total system jitter $\sigma^{(\mathrm{sys})}_t = 405$ ps.
From the Gaussian fit we see that $\sigma_{\tau} = 396$ ps, which is consistent with the systematic jitter $\sigma^{(\mathrm{sys})}_t$ within the fitting errors.
In fact, the correlation width is characteristic of the system and is relatively unchanged regardless of the channel conditions.
Therefore, the width of the correlation feature can be used as a test to determine if the algorithm was successful.
For example, if QTT and the peak finding algorithm obtain a noise peak, then the width would likely be much narrower than a true peak, and the erroneous $\tau$ could be disregarded.
We use this technique in finding the probability of success in Sec.~\ref{sec:Prob Success}.

\section{Simulation}\label{sec: Sim}

\subsection{Clock Offset}\label{sec:ClockOffset}
In principle, the performance of QTT is unaffected by the value of the clock offset, assuming that the range of the correlation histogram is large enough to include it.
However, in practice, measuring large clock offsets could require a prohibitively large amount of computation time unless treated properly.
For example, measuring a 1-second clock offset with 100-ps bins would require the correlation histogram to have at least $10^{10}$ bins. 
Therefore, in practice one divides large clock offsets into coarse and fine components \cite{valencia2004distant}, and if available one can use GPS or estimates of the propagation time to narrow the searching space.
Since we are interested in the fundamental performance of the QTT algorithm, we consider only the fine component of the clock offset $\tau$, and without loss of generality, we let $\tau=0$.

\subsection{Clock Drift}\label{sec: clock Stability}
The observed frequency drift between Alice and Bob's clocks $\Delta U$ can be caused by clock drift or relative motion. 
For systems with large clock drift $\Delta U$, it is important to model $\Delta U$ and preemptively subtract it from the measured time series using Eq.~\ref{eq: Coin Track}.
For example, it is common practice for GPS and laser-communication systems to model and remove the effects of Doppler shift, gravitational frequency shift, and other effects in order to prevent large timing errors \cite{ashby2002relativity}. 
Consequently, we omit the effects due to relative motion.
In other words, we assume course clock drift correction has been performed and study the performance of QTT subject to the clock drift $\Delta U$ contributions that are inherent to the QTT system.
This includes modeling errors and frequency jitter, which we assume to be a Gaussian distributed random variable with zero mean and standard deviation $\sigma_U$.
We discuss how the remaining $\Delta U$ contributions can negatively impact performance and limit the acquisition time $T_{\mathrm{a}}$ in Appendix~\ref{sec:AppendixClockStability}.
We consider three different clock stabilities: Rubidium frequency standards (RbFS), Cesium frequency standards (CsFS), and perfectly stable clocks (that is, $\sigma_U$ = 0). 
For the RbFS case, we performed QTT in our testbed and measured $\Delta U = 3.4 \times 10^{-10}$ and frequency jitter $\sigma_U = 3 \times 10^{-12}$.
For the CsFS case, we used the same $\Delta U$ but $\sigma_U = 5 \times 10^{-13}$ based on manufacture specifications \cite{MicrochipCsFS,everythingRF}.

\subsection{Heralding Efficiency}
We investigate the performance of QTT considering sources with different heralding efficiencies.
This is a pivotal consideration, because unlike sky noise photons that can be filtered with tighter spectral and spatial filtering, the photon source itself produces noise photons perfectly in band with probability $1-\eta_{\mathrm{herald}}$. 
We consider four different CW sources with heralding efficiencies that range from readily available commercial-off-the-shelf units to specialized one-off devices, namely $20$, $40$, $60$, and $80\%$, respectively.

\subsection{Photon Detection Statistics} \label{sec:DetStat}
\subsubsection{Photon Source}
Photon-pair sources have different observed statistics depending on their spatio-temporal mode structure and the temporal resolution of the detection system.
For example, theory predicts that a single-mode biphoton source, when generalized to a two-mode squeezed vacuum (TMSV) state, has thermal statistics.
However, this is only observed if the detection system has resolution finer than the coherence time of the photons, otherwise Poissonian statistics are observed \cite{PhysRev.131.2766,rockower1989self,schneider2018simulating}.
In the former case, it may be necessary to simulate thermal statistics to determine if there is a significant effect on the QTT algorithm.
In this simulation, regardless of whether the source is to be considered single or multimode, the system jitter is much larger than the few pico-second coherence time of the photons from the biphoton sources assumed here.
Therefore, it is sufficient to model the photon detection streams with Poisson statistics as we describe in the next section.
\subsubsection{Attenuation}
The phenomenon of a fluctuating atmospheric transmission, characterized by the probability distribution of the transmission coefficient (PDTC), simply modulates the temporal detection signal depending on the atmospheric conditions and properties of the optical receiver system \cite{semenov2009quantum}.  
However, Eq.~\ref{eq:TimeDiff} indicates that the QTT correlation histogram is sensitive to the difference of detection times at Alice and Bob, but is insensitive to where the pairs are located in the acquisition time $T_{\mathrm{a}}$.
Consequently, modulation of the pair arrival rate during the acquisition time $T_{\mathrm{a}}$ should not impact QTT.
Thus, it is sufficient to model the channel attenuation during the acquisition time $T_{\mathrm{a}}$ by a mean value.

\subsection{Simulation Methods}
We assume a CW biphoton source with a 2 Mcps pair rate $R$ and a 1 second acquisition time $T_{\mathrm{a}}$.
We assume the source generates photon pairs that obey Poisson statistics, which is equivalent to distributing the biphoton time tags randomly in $T_{\mathrm{a}}$ according to a uniform probability distribution.
Therefore, each of the Alice and Bob time series are generated in the following way.
First, we create an array of 2 million uniformly distributed random times between 0 and 1 second. 
The array is then sorted from smallest to largest time, becoming the unattenuated Alice time series.
An identical copy is made for Bob, which would be shifted appropriately relative to the Alice time series for cases where the clock offset $\tau$ is non-zero.

The effects of the total systematic timing jitter are now included by adding a Gaussian distributed random variable with standard deviation $\sigma_{t_\mathrm{A}}$ and $\sigma_{t_\mathrm{B}}$ to each time tag of Alice's and Bob's time series, respectively.
Unless otherwise noted, we assume SPAD detectors and time taggers with timing jitter $\sigma^{(\mathrm{det})}_t=287$ ps and $\sigma^{(\mathrm{tt})}_t=4$ ps, respectively.
These values were chosen to match the typical timing jitters we observe with Excelitas SPCM-AQRH single-photon counting modules and PicoQuant HydraHarp 400.

Next, time tags are randomly removed from Alice's and Bob's time series equivalent to their respective channel attenuation and the heralding efficiency of the source $\eta_{\mathrm{herald}}$.
Alice's channel attenuation is set to the product of the detection efficiency $\eta_{\mathrm{det}}=0.6$ and the losses due to the transmission of the spectral filter $\eta_{\mathrm{spec}}=0.9$.
Bob's channel attenuation and background noise photons are set to span -10 to -50 dB and 0 to 800 kcps, respectively.  
We will show that these values include conditions commensurate with a daytime space-to-Earth downlink.

Next, we include the effect of dead time, which can reduce the probability of Alice and Bob detecting true coincidences as the heralding efficiency or the background rate at Bob increases. 
We chose an 84 ns dead time to match the total systematic dead time that we have observed in our hardware.
It is worthwhile to note that single photon detectors can be paralyzable or nonparalyzable, which can complicate the simulation procedure.
Nonparalyzable detectors are not affected by a photon that arrives during the dead time, whereas paralyzable detectors reset if a photon arrives during the dead time \cite{knoll2010radiation}.
In this simulation, we assume a worst case scenario by modeling the paralyzable case and removing all time tags from Alice's and Bob's time series that are within the dead time of another time tag.
However, we note that for Bob's channel conditions assumed here, we find that the paralyzable and nonparalyzable cases give similar results.
Lastly, we find that the after pulsing probability is much less that $1\%$ by the time the dead time has elapsed \cite{ziarkash2018comparative}.
As a consequence, after pulsing does not contribute significantly to the QTT performance over the parameter space of interest in this work and it is omitted in this simulation.

This concludes the preparation of Alice's and Bob's simulated time series and the QTT algorithm can now be performed. 
Since the clock offset $\tau$ is determined based on the mean of the Gaussian fit of the correlation peak, it is important to have a reliable fit.
This translates into choosing the correlation bin size $T_{\mathrm{bin}}$ small enough to resolve the correlation peak but not so small that noise fluctuations effect the fit.
We find that 100 ps and 10 ps correlation bin widths are compatible with SPAD and superconducting nanowire single-photon (SNSPD) detectors, respectively.

\subsection{Probability of Success} \label{sec:Prob Success}
\begin{figure*}[]
	\includegraphics[width=0.9\columnwidth]{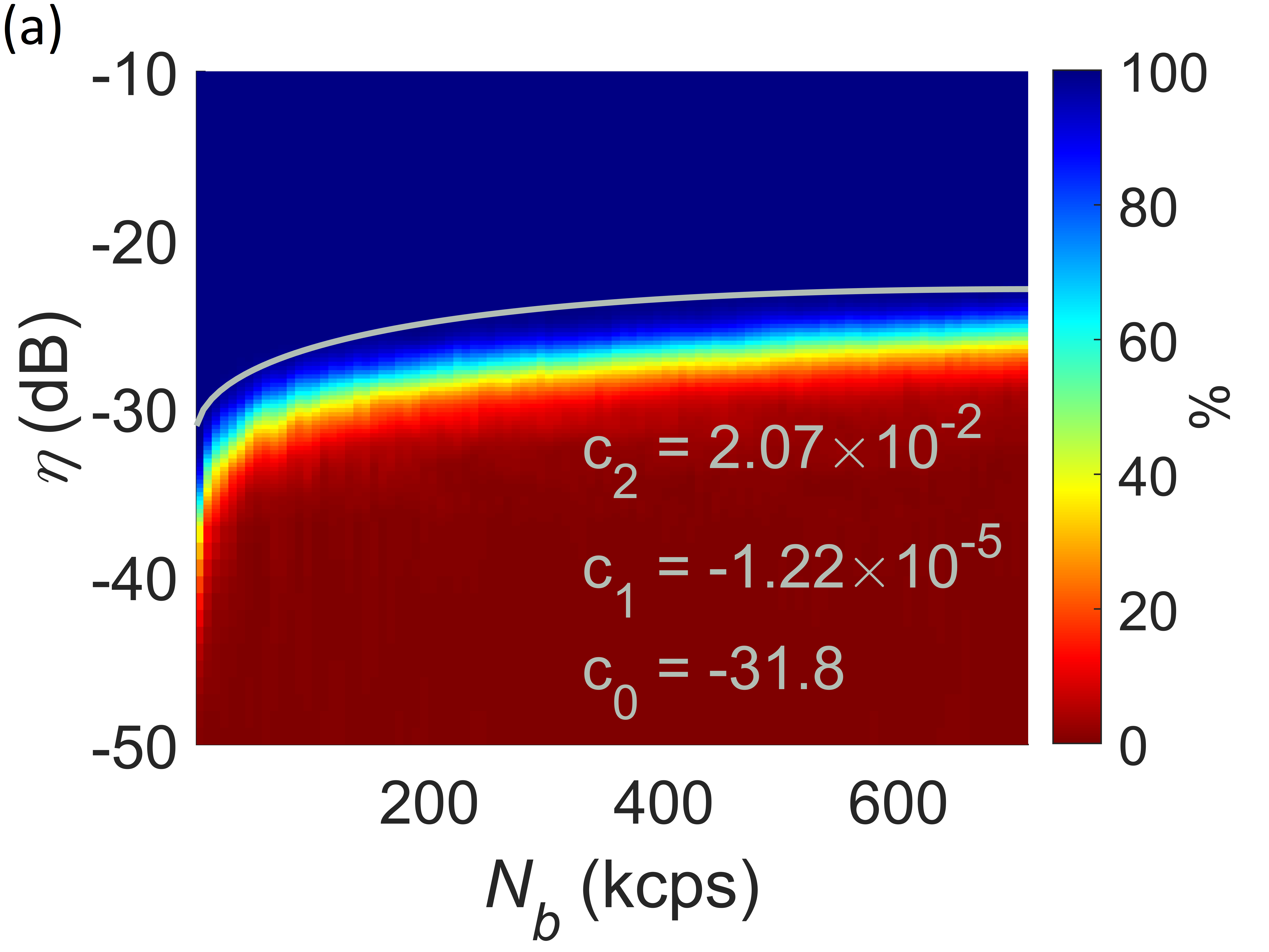}%
	\vspace{0.0002cm}
	\includegraphics[width=0.9\columnwidth]{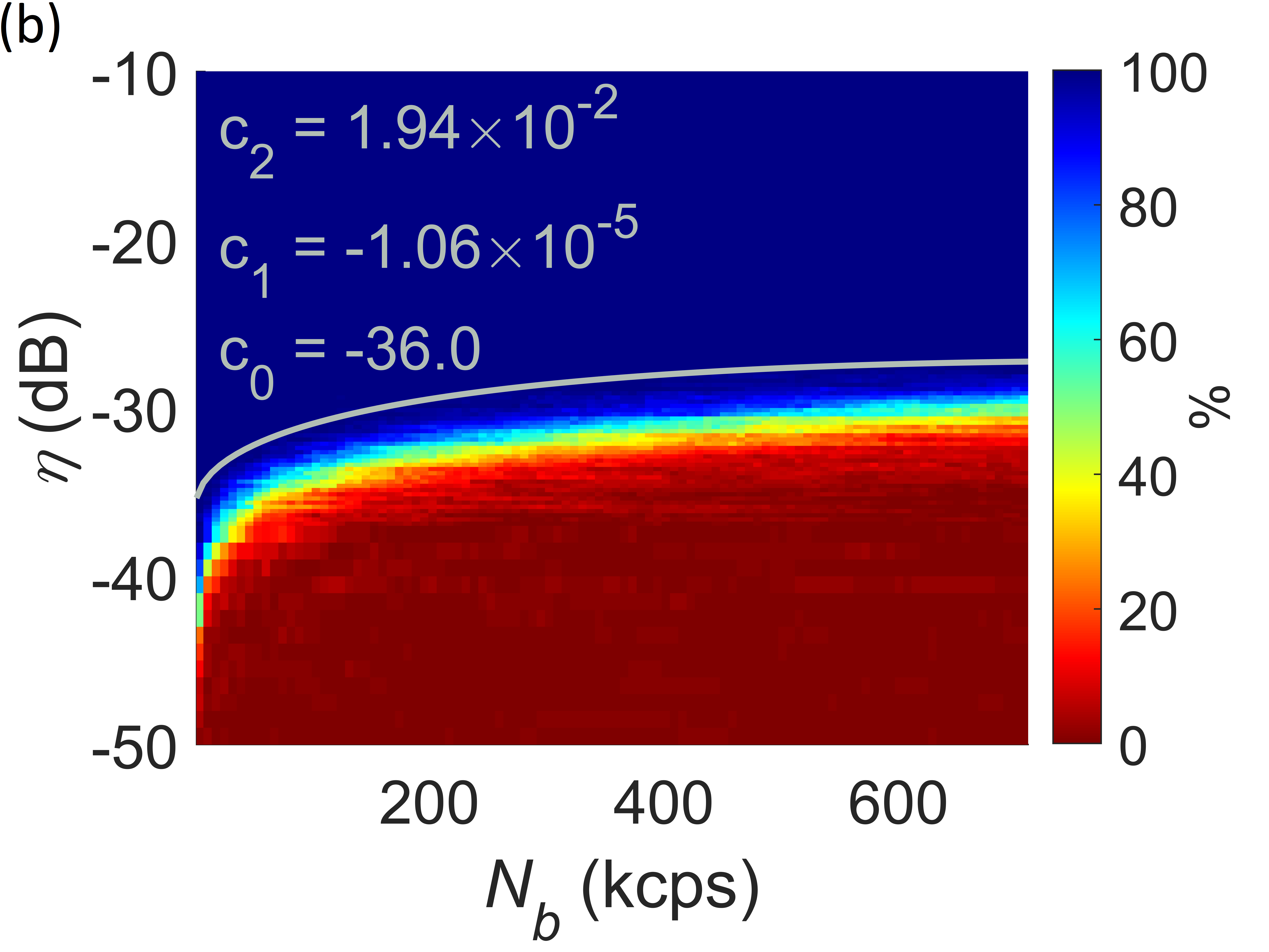}\\
	\vspace{0.0002cm}
	\includegraphics[width=0.9\columnwidth]{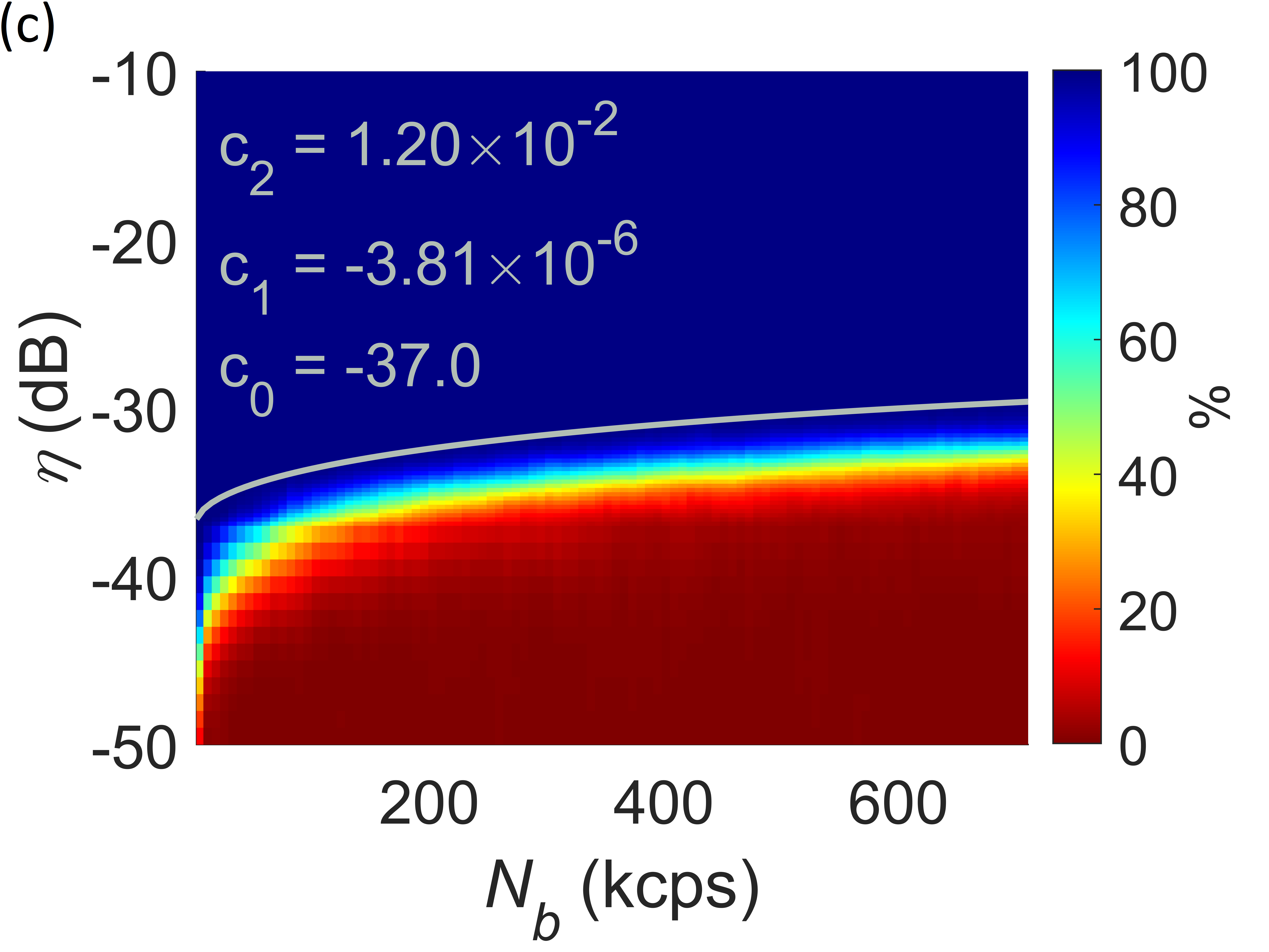}%
	\vspace{0.0002cm}
	\includegraphics[width=0.9\columnwidth]{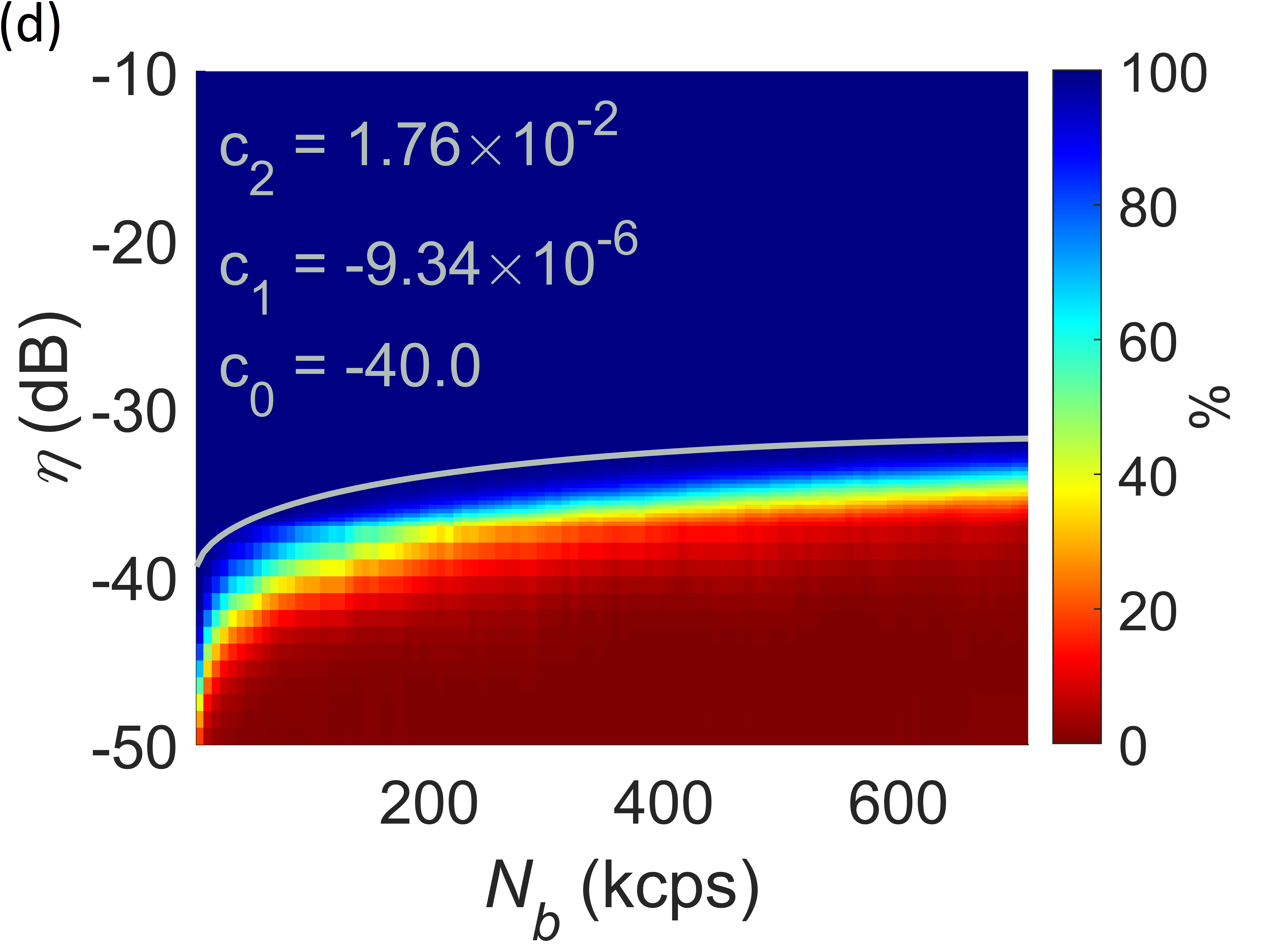}\\ 
	\caption{\label{fig:prob_succ}
Density plots giving the probability of successful QTT as a function of Bob's channel attenuation $\eta_{\mathrm{ch}}$ and background counts $N_b$.
(a)--(d)  Gives the results for $\eta_{\mathrm{herald}}$ equal to 20, 40, 60, and 80$\%$, respectively. The gray line traces out the threshold channel attenuation that can be achieved with $99\%$ probability of success and the inset gives the fit parameters corresponding to Eq.~\ref{eq:ThrAtt}, where $c_0$ is the fit parameter for the y-intercept.
	}
\end{figure*}
In this section, we span a 2D parameter space of mean channel attenuation and background-noise photons.
We repeat each channel condition 100 times to find statistically relevant quantities.  
For each channel condition we examine the probability of success, which is the number of times the QTT algorithm correctly identified the clock offset in the 100 trials.
Thus, the probability of success is the probability that a single instance of the QTT algorithm will return the correct clock offset.  
We will refer to this as the ``single-shot" probability of success. 

In Figs.~\ref{fig:prob_succ}~(a)--(d) we show the probability of success as a function of attenuation and the number of noise photons for sources with 20, 40, 60, and 80$\%$ heralding efficiencies, respectively.
This shows that the QTT algorithm is highly robust to noise photons and is much more susceptible to channel loss as seen by the abrupt drop in performance with increasing channel attenuation $\eta_{\mathrm{ch}}$. 
The gray line is a fit to the 99$\%$ probability of success threshold using the framework derived in Appendix~\ref{sec:Analytic Model}.

\subsection{Threshold Attenuation}
\begin{figure}[t!]
	\includegraphics[width=0.9\columnwidth]{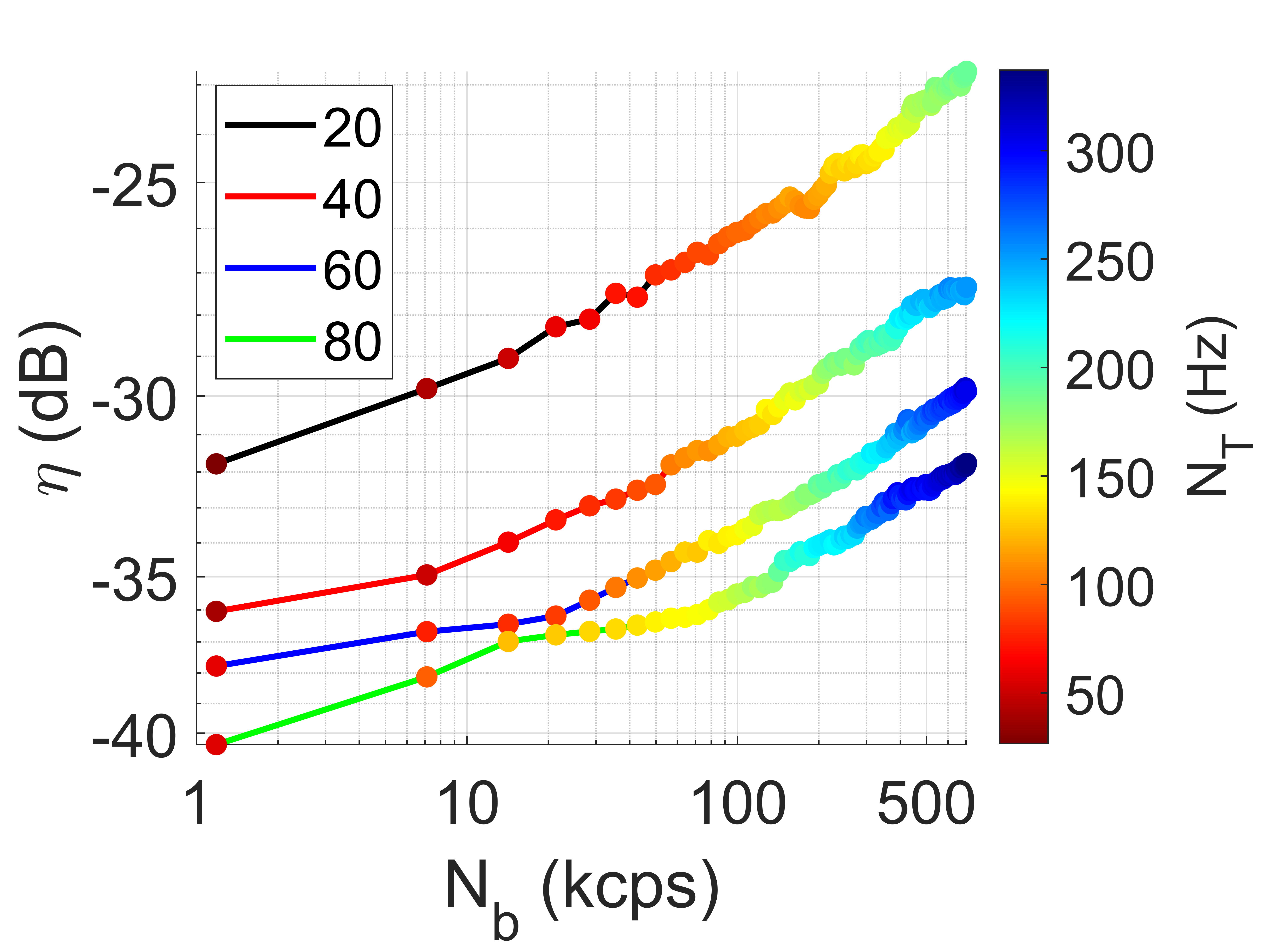}\\
	\caption{\label{fig:prob_succ_thr}
Curves giving the threshold attenuation $\eta_{\mathrm{th}}$ that maintains $\geq$ $99 \%$ probability of success as the number of sky background photons increases. The black, red, blue, and green curves correspond to 20, 40, 60, and 80$\%$ heralding efficiency, respectively. The marker colors correspond to the number of true coincidences for each of the channel conditions.
	}
\end{figure}

In Fig.~\ref{fig:prob_succ_thr} we impose a $99\%$ single-shot probability of success threshold and find the corresponding channel attenuation $\eta_{\mathrm{th}}$ for 20, 40, 60, and 80$\%$ heralding efficiency.
For each heralding efficiency, the marker color indicates the number of true coincidences $N_{\mathrm{T}}$.
As expected, we see that more attenuation $\eta_{\mathrm{th}}$ can be managed as the heralding efficiency increases.
This is simply because higher heralding efficiencies correspond to larger true coincidence rates $N_{\mathrm{T}}$ and therefore larger coincidence peaks.
We see the QTT algorithm achieves $99\%$ success probability despite 100-kcps scale background-noise photons $N_b$, compared to only a few hundred cps received true coincidences $N_{\mathrm{T}}$.
This is because the sky noise photons are uncorrelated with the true coincidences and their contribution is spread out uniformly over the entire histogram range.

\subsection{SEM}
\begin{figure}[t!]
	\includegraphics[width=0.9\linewidth]{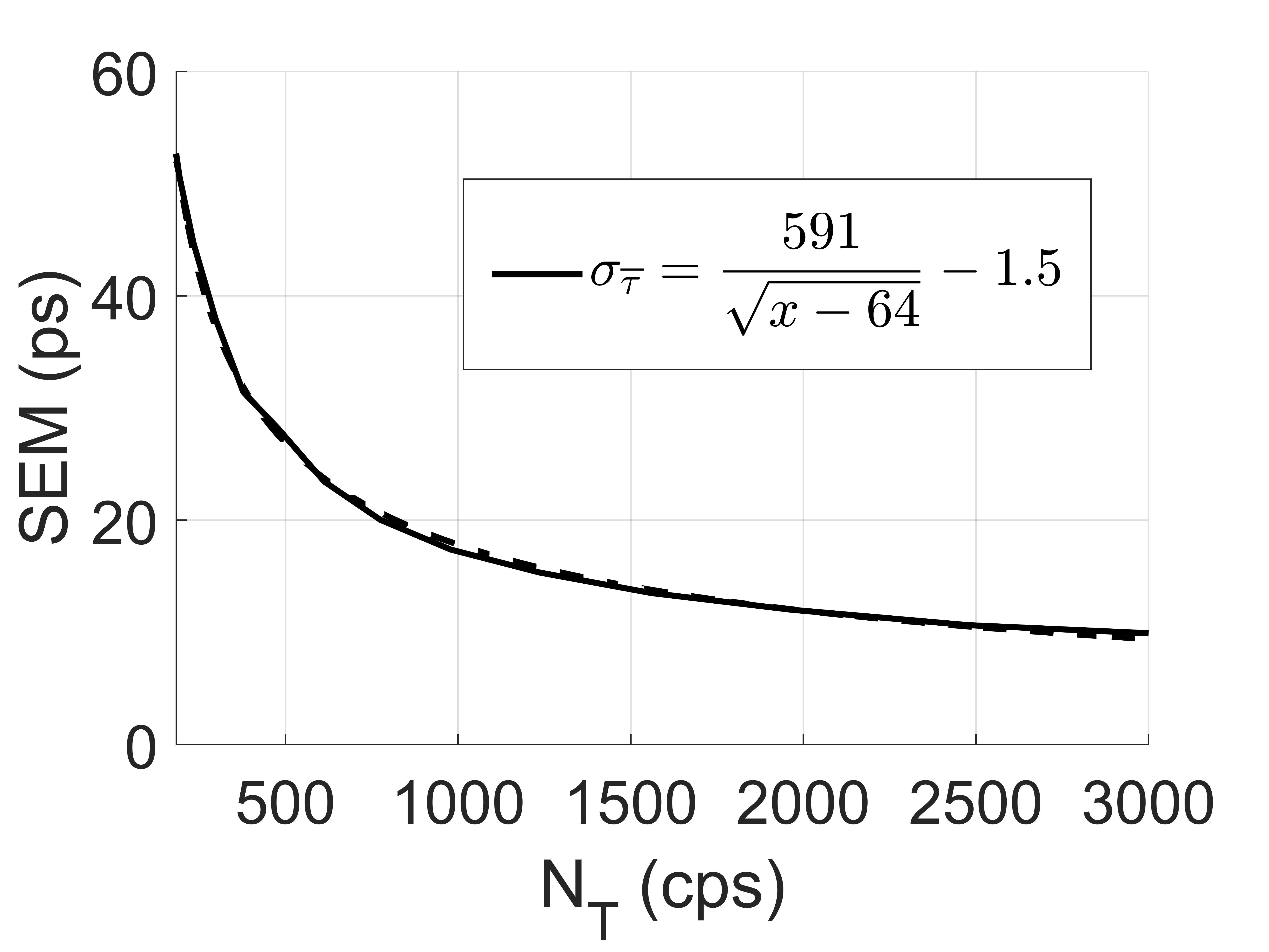}		
	\caption{The SEM clock offset as a function of the number of ``true'' coincidences $N_{\mathrm{T}}$ (that is, with accidentals subtracted), where the number of background photons are set to $N_b = 2$ Mcps.   On the black curve, each SEM is calculated by taking the standard deviation of the clock offsets measured from a Monte Carlo simulation with 1000 trials.  The dashed curve is the fit and the fit parameters are given in the plot legend.}
	\label{fig:SEM}
\end{figure}

The QTT error should converge to Eq.~\ref{eq: SEM equ} in the limit of large number of true coincidences (i.e. large sample size).
To measure the QTT error, i.e. the \textit{true} SEM $\sigma_{\overline{\tau}}$, we perform the following Monte Carlo simulation.
First, we measure the sampling distribution by simulating 1000 independent clock offset measurements, where we fixed the number of background photons at a daytime rate of $N_b\sim$2 Mcps.
We then measure the SEM by calculating the standard deviation of the sampling distribution.
Lastly, we repeat these steps over a range of true coincidences $N_{\mathrm{T}}$.
Figure~\ref{fig:SEM} gives the true SEM of the QTT algorithm, where, given the $200$ Mcps pair rate and $40\%$ heralding efficiency of the source, $N_{\mathrm{T}}\sim 800$ represents the expected number of true coincidences during a daytime space-to-Earth downlink. 
The fit in the inset reveals the characteristic $1/\sqrt{N_{\mathrm{T}}}$ dependence of the SEM, but the numerator of the fit, 591 ps, is $\sim$1.5 times larger than the SEM predicted by Eq. \ref{eq: SEM equ}.
If one increases $N_{\mathrm{T}}$ to large values, while also decreasing the correlation bin width $T_{\mathrm{bin}}$ in proportion, then the difference between the SEM calculations will tend to decrease, but only up to a limit.
This is most likely related to errors incurred when fitting the correlation peak. 
Nevertheless, Fig. ~\ref{fig:SEM} shows that the timing precision of a single-shot clock offset measurement $\tau$ is proportional to $1/\sqrt{N_{\mathrm{T}}}$, and the proportionality can be modeled directly depending on the source and channel conditions.

\subsection{Allan Deviation}
\label{sec: Allan Dev}
In this section we discuss the clock stability of QTT.
The Allan deviation $\sigma_\mathrm{y}$ is a standard method to characterize the stability of a clock system.
When the Allan deviation is plotted on a log-log scale, the slope of the curve indicates the type of noise in the system.
Thus far, we have focused on the performance of independent, single-shot QTT measurements.
In order to model the stability, we simulate continuous streams of successive QTT measurements under the influence of drifting \textit{and} jittering local clocks.
The effect of clock drift and relative motion sets the mean $\Delta U$, which we include as a modification of Bob's time series $t_B$ according to Eq.~\ref{eq: freq drift}. 
The jitter is modeled as a Gaussian random variable of standard deviation $\sigma_{U}$, which varies depending on the stability of the local clocks.
We consider daytime and nighttime scenarios and investigate the effect of using RbFS, CsFS, or perfect clocks paired with SPADs or SNSPDs on the Allan deviation.
\begin{figure*}[t!]
	\includegraphics[width=0.95\linewidth]{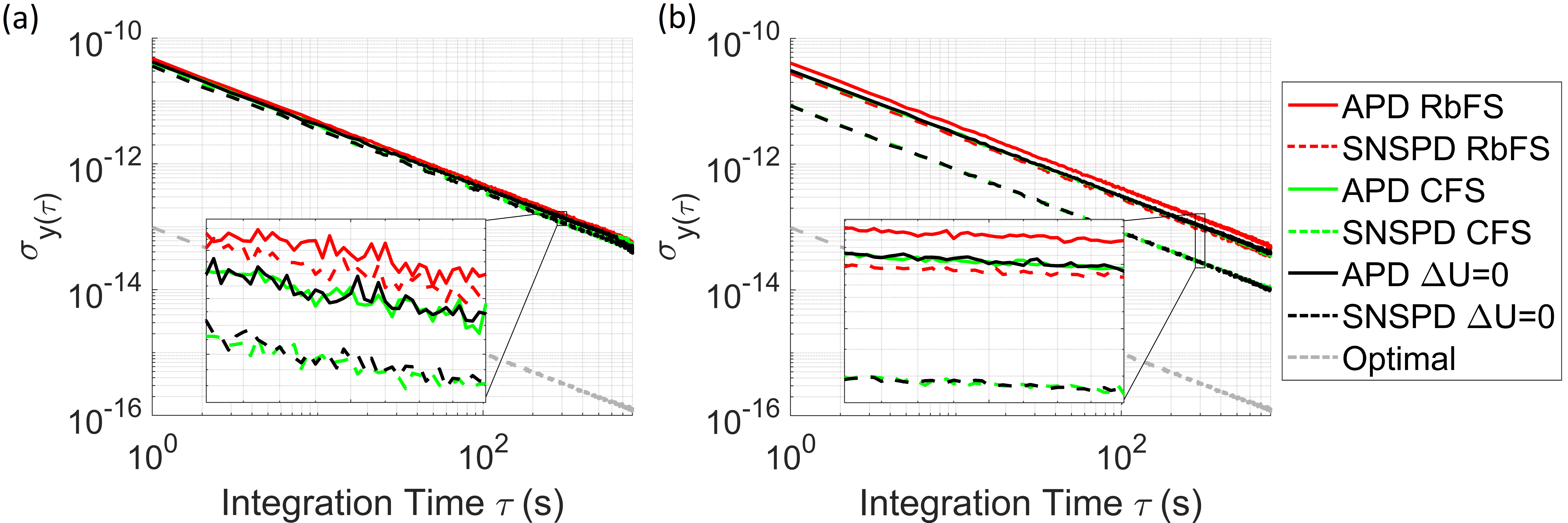}
	\caption{The overlapping Allan deviation $\sigma_y$ for channel attenuation $\eta_{\mathrm{ch}} = -23$ dB and daytime (a) $N_b = 2.14$ Mcps and (b) nighttime $N_b = 100$ kcps sky-background rates. The red, green, and black curves correspond to RbFS, CsFS, and perfect clocks respectively. The solid curves indicate SPADs whereas the dashed curves indicate SNSPDs. The gray line is representative of a lower bound based on benign channel conditions and SOTA clock components: $\eta_{\mathrm{ch}} = -5.7$ dB, $N_b=300$, $\Delta U = 3\times10^{-17}$, and $\sigma_U=2\times10^{-16}$. The inset shows the relative timing stability for the different cases over the integration range from 280 to 320 seconds.}
	\label{fig:allan_deviation}
\end{figure*}

Figure~\ref{fig:allan_deviation} (a) shows the overlapping Allan deviation \cite{riley2008handbook} with channel conditions commensurate with a daytime space-to-Earth downlink utilizing adaptive optics (AO) wavefront correction. 
The red, green, and black curves correspond to RbFS, CsFS, and perfect clocks, respectively.
The solid curves indicate SPADs whereas the dashed curves indicate SNSPDs.
The gray line is representative of a lower bound based on benign channel conditions and state-of-the-art (SOTA) clock components: $\eta_{\mathrm{ch}} = -5.7$ dB, $N_b=300$, $\Delta U = 3\times10^{-17}$, and $\sigma_U=2\times10^{-16}$.
The negative $1$ slope indicates white phase modulation noise, which is consistent with the noise of frequency standards operated in phase-locked control loops \cite{sullivan1990characterization, riley2008handbook}.
The inset shows the relative timing stability for the different cases over the range of integration time from 280 to 320 seconds.
As expected, there is an increase in stability as one uses detectors with less jitter and clocks with more stability.
Reduced detector jitter results in smaller correlation widths $\sigma_\tau$, which in turn causes smaller SEM $\sigma_{\overline{\tau}}$ and better stability.
Interestingly, when using either SPADs or SNSPDs, switching from CsFSs to ``perfect clocks"  does not improve the stability.
This shows that the detection jitter of SPADs or SNSPDs is the limiting factor when paired with clocks that are at least as stable as CsFSs.

Figure~\ref{fig:allan_deviation} (b) shows the overlapping Allan deviation with channel conditions commensurate with a nighttime downlink utilizing AO.
In this case the noise is reduced by a factor of $\sim$21 and the increased signal-to-noise permits the scenario utilizing RbFS and SNSPDs (dashed red) to be comparable with CsFS and SPAD case (solid green). 
Overall, this shows how one can explore the tradespace of channel conditions and equipment specifications in order to meet a performance objective. 

\section{Relevance to Telecom Channels}\label{sec:FiberLength}
\begin{figure}[t!]
	\includegraphics[width=0.9\linewidth]{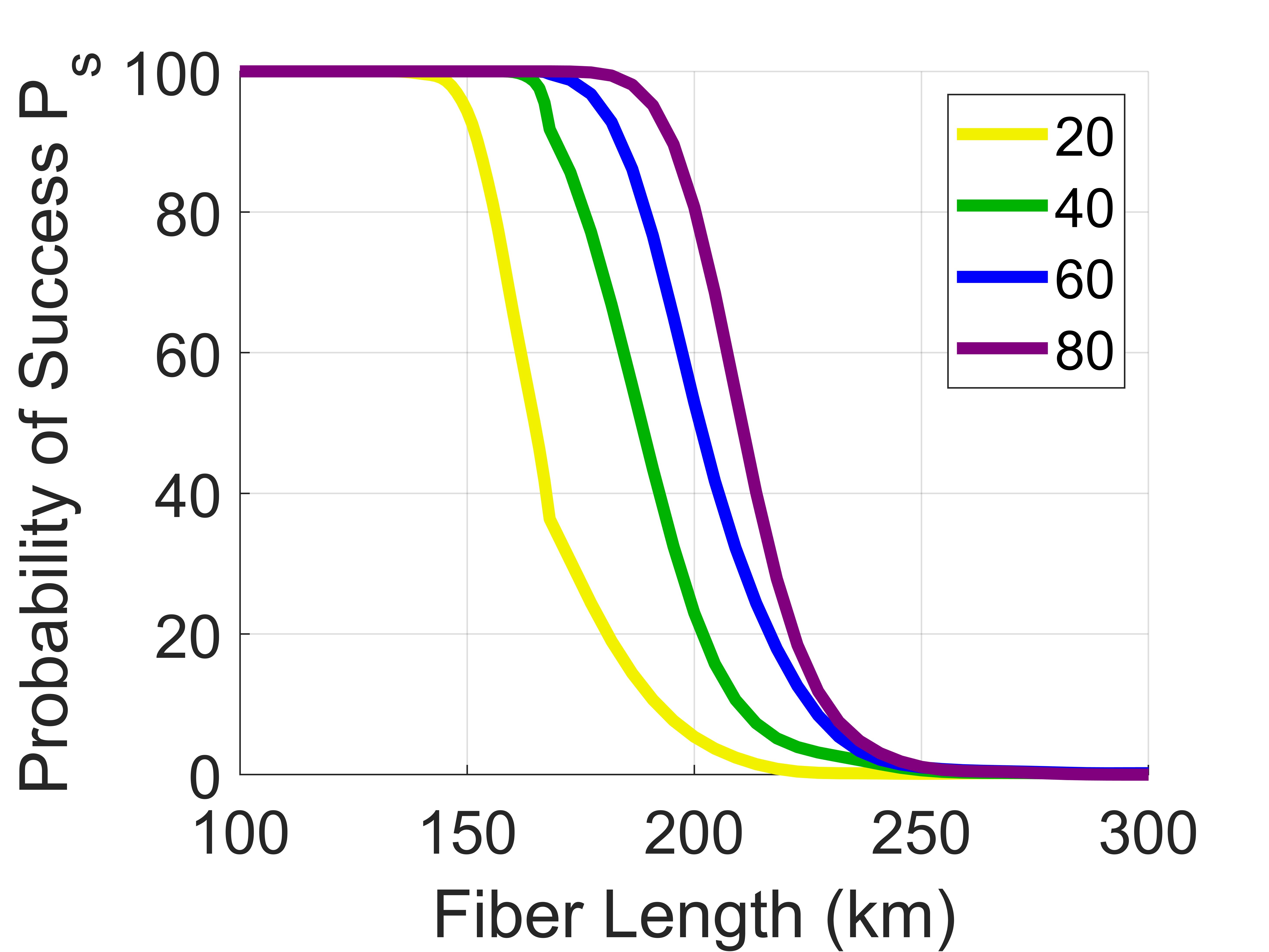}		
	\caption{The probability of success $P_s$ as a function of the total length of fiber between Alice and Bob for a biphoton source with a pair rate of 2 Mcps and 20, 40, 60, and 80 $\%$ heralding efficiency and $N_b \approx 1$ kcps background rate.}
	\label{fig:Fiber_Channel}
\end{figure}

To demonstrate the applicability of our QTT algorithm and simulation, we first apply our results to a telecom-fiber channel.
We do this by converting the channel attenuation $\eta_{\mathrm{ch}}$ to fiber length according to $\eta=-\alpha L$, where $L$ is the total fiber length in kilometers and $\alpha = 0.22$ dB/km is the attenuation coefficient at 1550 nm.
The result with $N_b\approx1$ kcps is shown in Fig. \ref{fig:Fiber_Channel}, where the probability of success $P_s$ is plotted as a function of the total length of fiber between Alice and Bob.
As expected, increasing the source heralding efficiency allows the QTT algorithm to perform better at longer fiber lengths.
Furthermore, it shows that the QTT algorithm achieves high probability of success up to a few hundred kilometers given the source rate and heralding efficiencies that we model.

\section{Relevance to Space-Earth Channels}\label{sec:HEMIS}
To further demonstrate the applicability of our QTT algorithm and simulation, we apply our results to channel conditions representative of a daytime space-to-Earth quantum downlink.
We assume a satellite in a 600-km circular orbit has a 15-cm transmit aperture and propagates 780-nm photons to a ground station with a 1-m receive aperture utilizing a 1-nm spectral filter. 
It is assumed that all the detectors in the system have efficiency $\eta_{\mathrm{det}}=0.6$, the spectral filters have efficiency $\eta_{\mathrm{spec}}=0.9$, and the ground station receiver has efficiency $\eta_{\mathrm{rec}}=0.5$.
The angle dependent atmospheric transmission efficiency $\eta_{\mathrm{trans}}$ and the background sky radiance $H_b$ are generated in MODTRAN for a high desert climate with urban aerosols.

In Ref.~\cite{lanning2021quantum} we establish a theoretical framework in which the receiver performance is modeled as a function of atmospheric phenomena and AO system parameters.
It uses scaling law equations to estimate the residual wavefront-phase error, after correction, and determine how the error inhibits the ability to transmit the signal light through a small spatial filter in the focal plane.
In summary, the Greenwood frequency $f_{\mathrm{G}}$ characterizes the rate at which the turbulence is changing, and the Fried coherence length $r_0$ characterizes the spatial scale of the turbulence. 
The tip, tilt, and higher-order correction of an AO system counteract the negative effects of these phenomena.
An AO system can be characterized by the closed-loop bandwidths of the system, the number of wavefront sensor subapertures, and the number of actuators making up the deformable mirror (DM) that applies the correction.
In the following subsections we are careful to present the residual errors that quantify the effect of these competing phenomena. 

We consider two conservative cases of optical receiver configuration.
First, we assume the receiver utilizes tracking alone, that is, only tip and tilt correction, with a tracking closed-loop bandwidth $f_\mathrm{TC} = 50$ Hz.
Secondly, we assume the receiver is configured with both tracking and higher-order AO correction utilizing $N_{\mathrm{act}} = 25$ mirror actuators and closed loop bandwidth $f_c = 100$ Hz. 
In both cases, we use the aforementioned framework to map the results of our simulation to a daytime sky hemisphere to show the performance of the QTT algorithm for different heralding efficiency sources.

\subsection{Tracking Only}\label{sec:HEMISwoAO}
In the case of utilizing tracking alone, one can estimate the residual wavefront-phase error $\sigma_{\phi}$ as a combination of the error from the higher-order structure of the signal light \cite{noll1976zernike} and the finite bandwidth of the tracking system \cite{hardy1998adaptive}
\begin{equation}\label{eq:RPEtrack}
\sigma_{\phi,\mathrm{CL}}^2 = 0.582 \Big( \dfrac{D_R}{r_0} \Big) ^{5/3} + \Big( \dfrac{\pi}{2} \dfrac{f_\mathrm{TG}}{f_{\mathrm{TC}}} \Big) ^{2}, 
\end{equation}
respectively, where $D_R$ is the diameter of the receiver aperture and $f_{\mathrm{TG}}$ is the tracking Greenwood frequency.
This residual error is inserted in place of the terms in the brackets of Eq.~A9 of Ref.~\cite{lanning2021quantum}, and the rest of the framework is unaltered, except for the zenith angle dependence of the atmospheric parameters \cite{hardy1998adaptive}.

\begin{figure}[]
	\includegraphics[width=0.9\columnwidth]{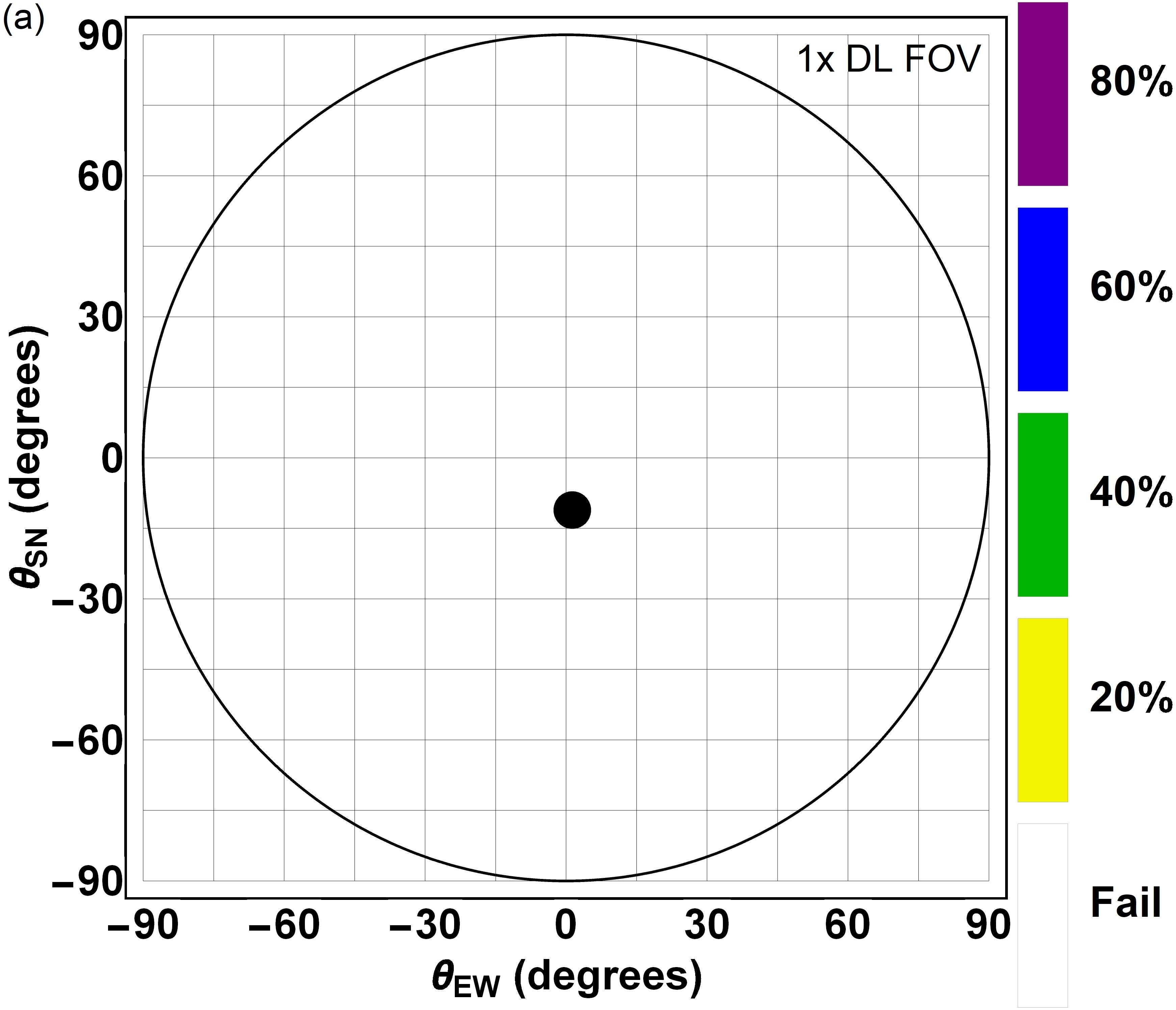}\\
	\vspace{0.0001cm} 
 	\includegraphics[width=0.9\columnwidth]{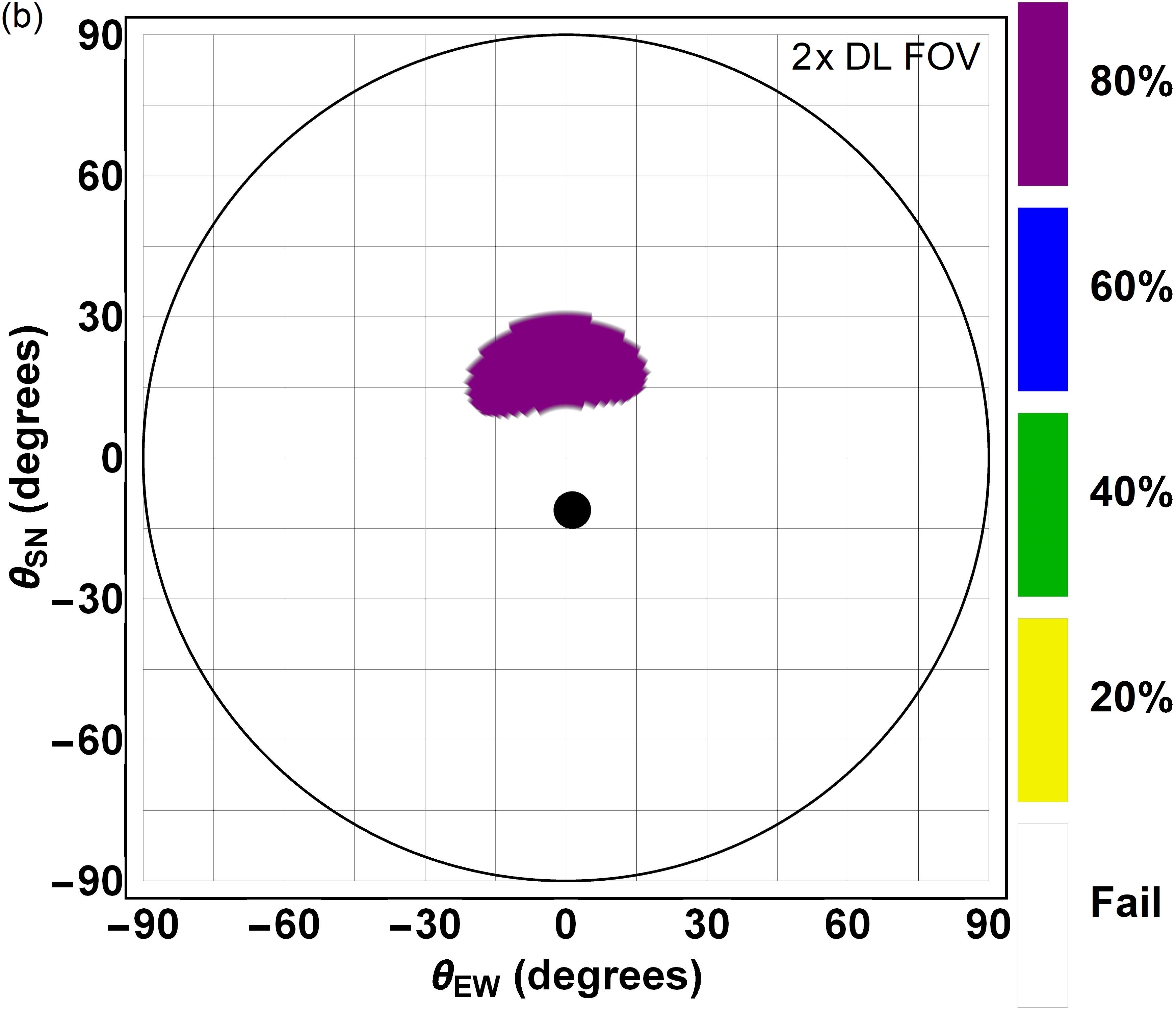}\\
 	\vspace{0.0001cm}
 	\includegraphics[width=0.9\columnwidth]{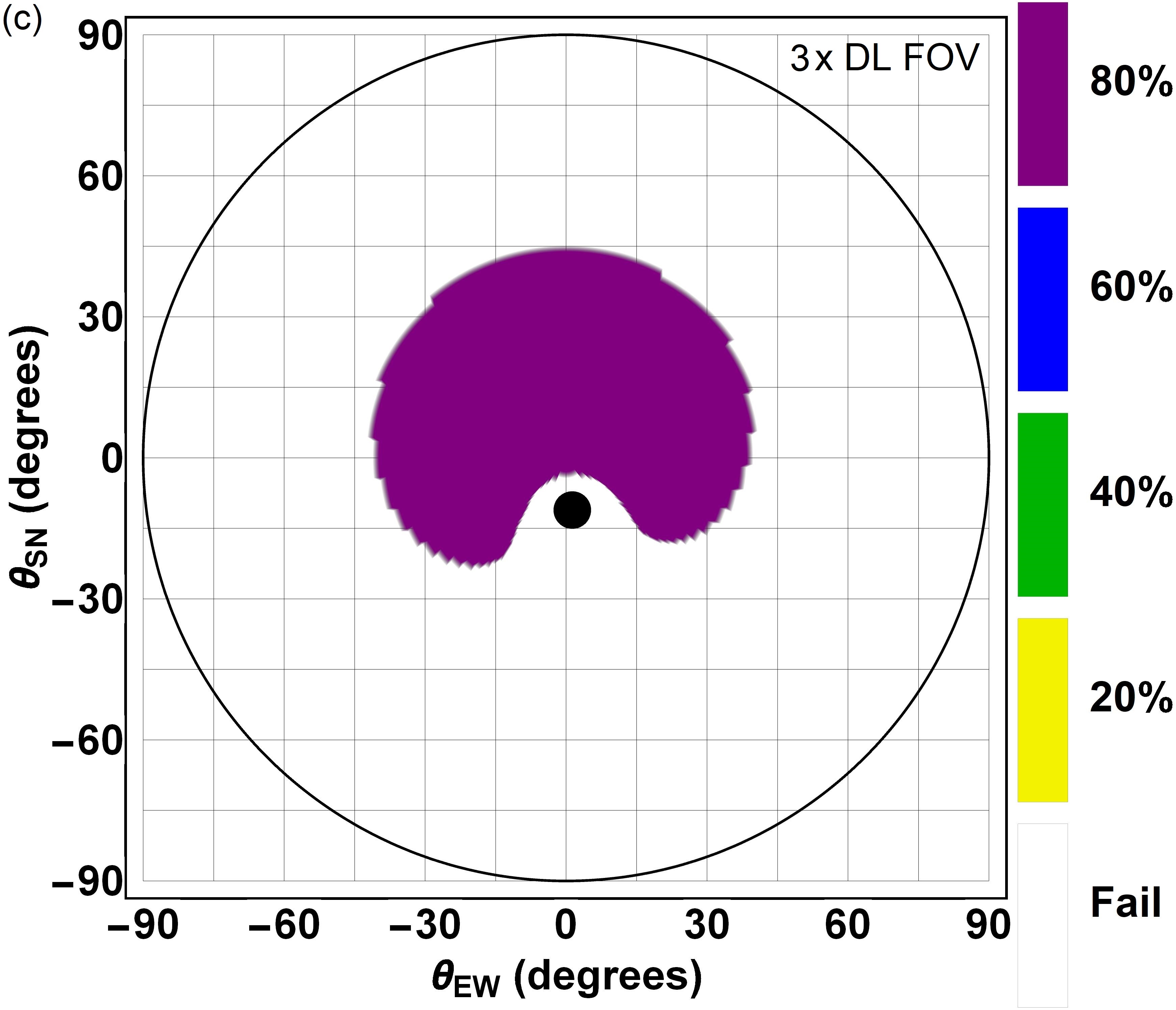}
	\caption{\label{fig:SUMMERnoAO}
Hemispherical plots of summer solstice with a 1$\times$HV$_{5/7}$ turbulence profile showing the regions where QTT succeeds with 99$\%$ probability for a receiver system with no higher-order AO (see Sec.~\ref{sec:HEMISwoAO} for more details). The legend gives the heralding efficiency of the biphoton source. In (a)--(c) we enlarge the FOV from 1$\times$ to 3$\times$ the diffraction limited FOV.  The black dot represents the sun location.
	}
\end{figure}

\begin{figure}[]
	\includegraphics[width=0.9\columnwidth]{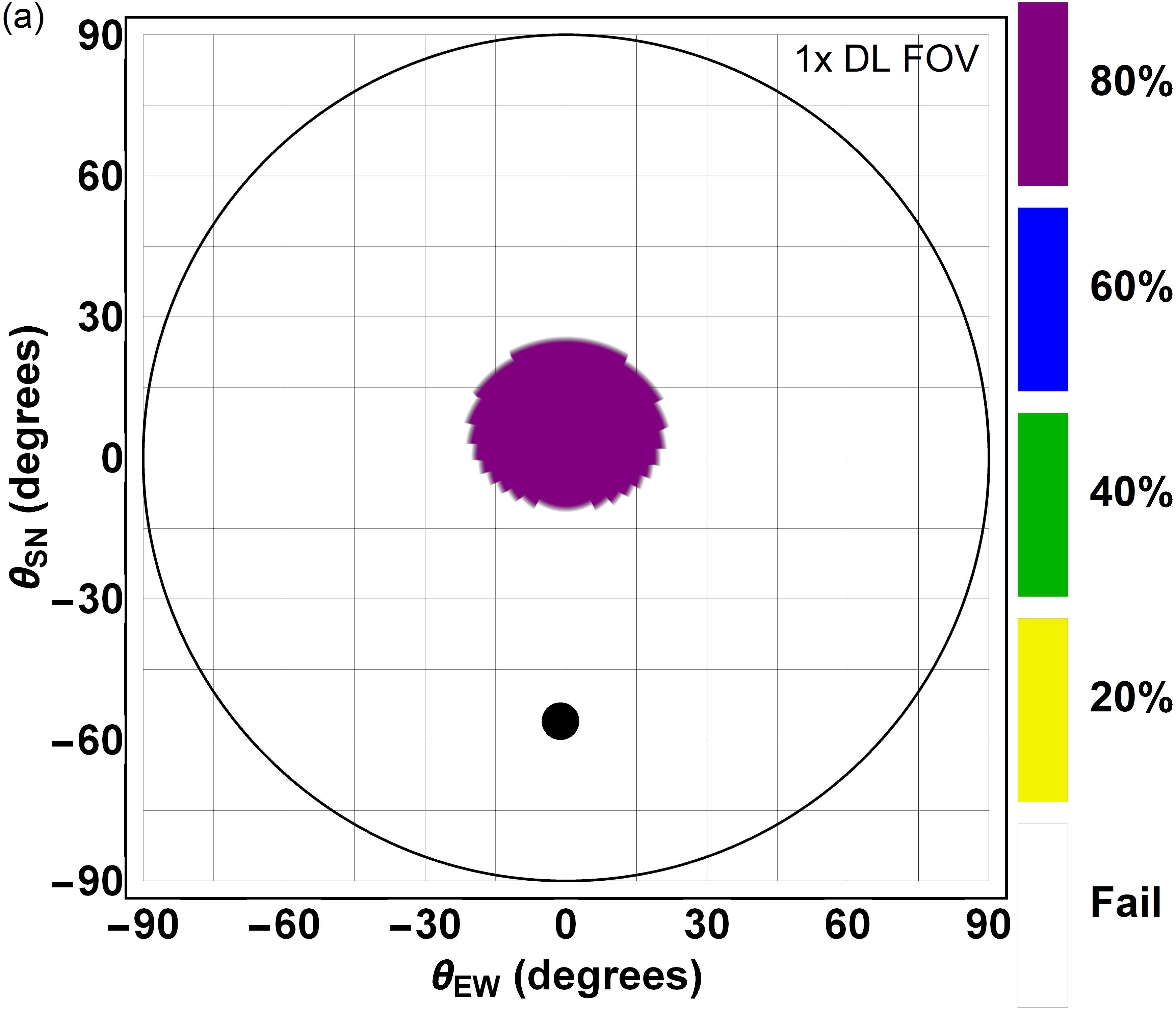}\\
	\vspace{0.0001cm} 
 	\includegraphics[width=0.9\columnwidth]{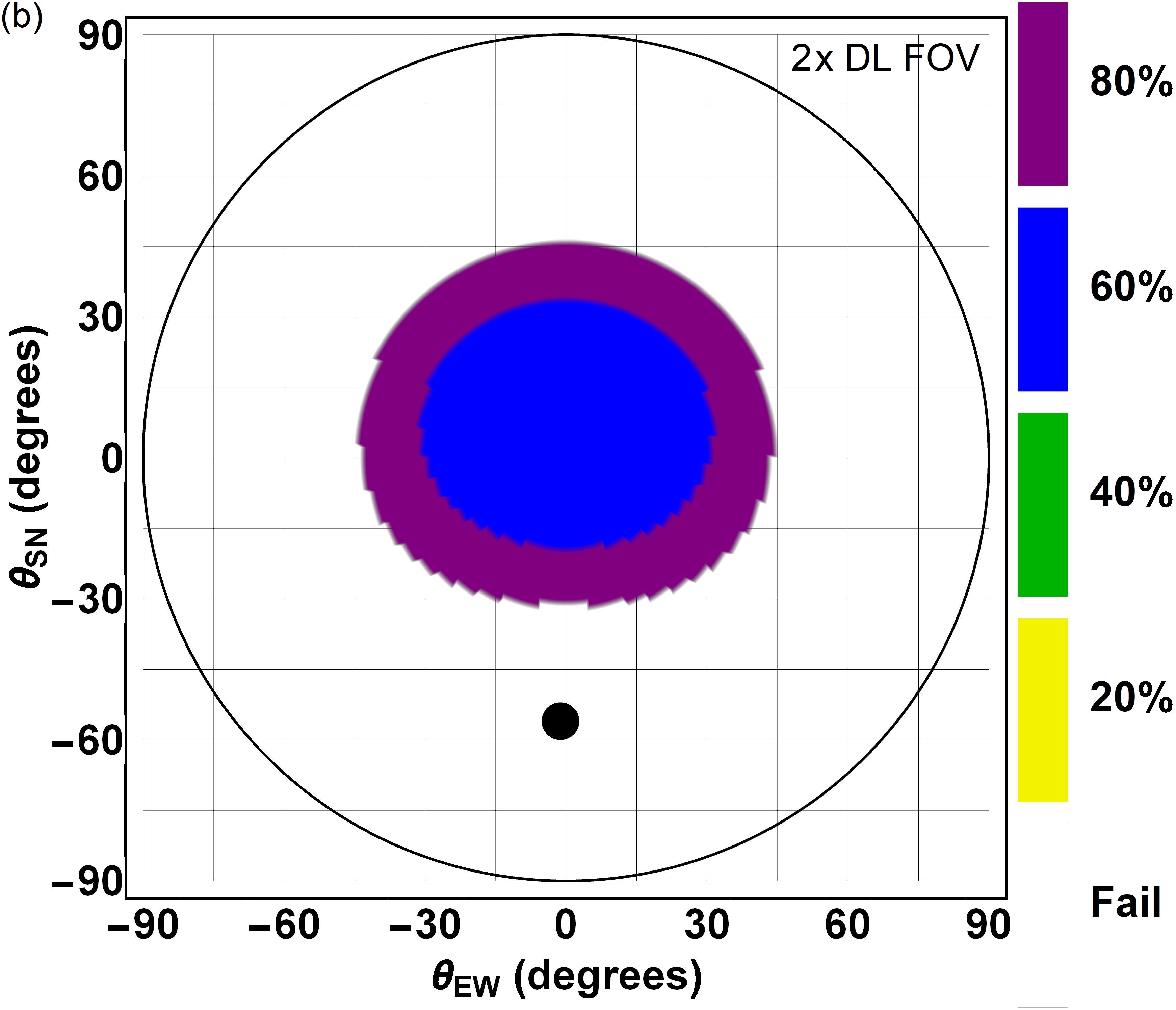}\\
 	\vspace{0.0001cm}
 	\includegraphics[width=0.9\columnwidth]{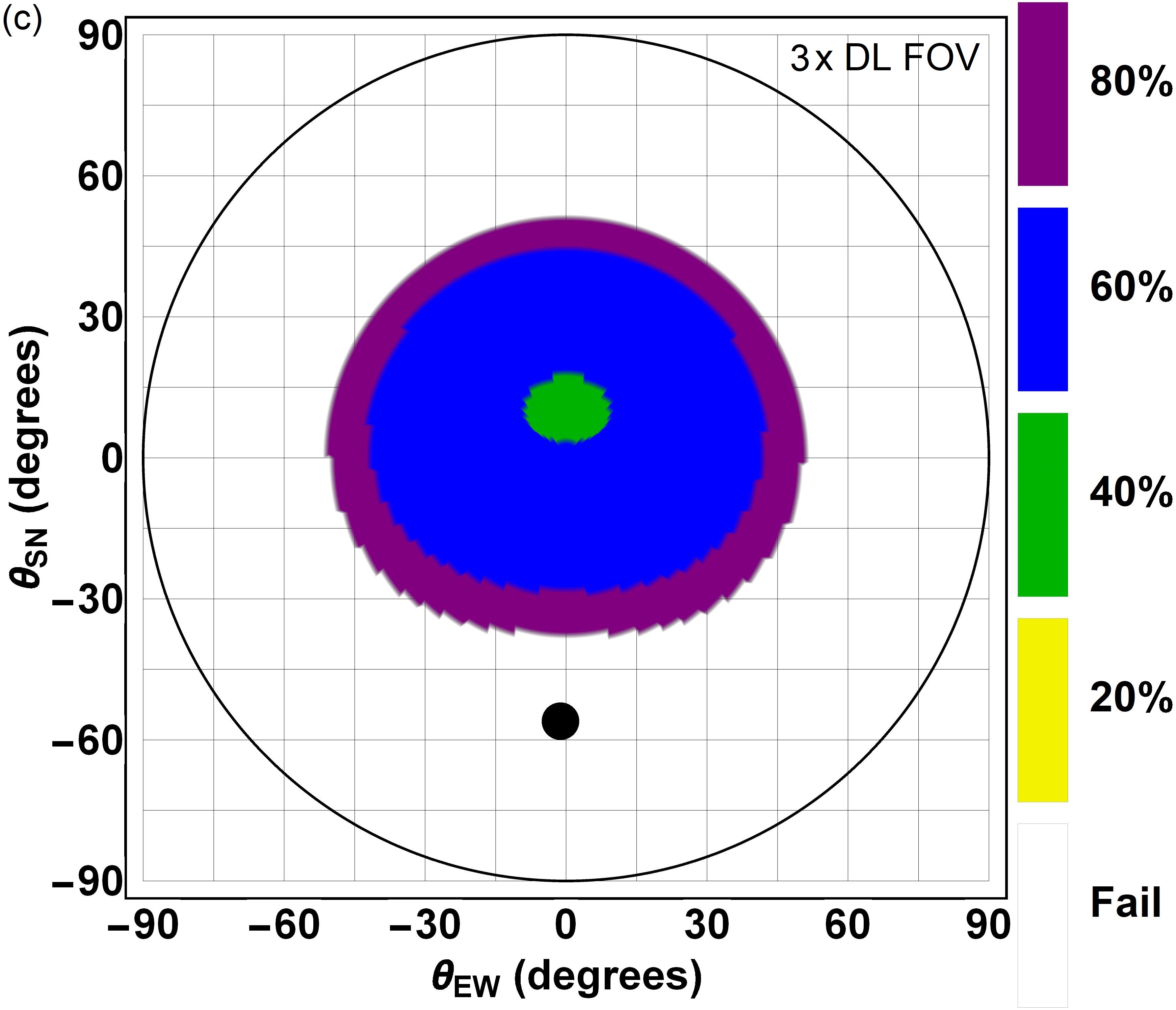}
	\caption{\label{fig:WINTERnoAO}
Hemispherical plots of winter solstice with a 1$\times$HV$_{5/7}$ turbulence profile showing the regions where QTT succeeds with 99$\%$ probability for a receiver system with no higher-order AO (see Sec.~\ref{sec:HEMISwoAO} for more details). The legend gives the heralding efficiency of the biphoton source. In (a)--(c) we enlarge the FOV from 1$\times$ to 3$\times$ the diffraction limited FOV.  The black dot represents the sun location.  
	}
\end{figure}

Using this map, we are able to generate hemispherical plots for two sun positions and varying turbulence strengths.
The plots show the regions where QTT succeeds with 99$\%$ probability, color coded by the heralding efficiency of the biphoton source.
Given the aperture sizes and channel conditions assumed, without higher-order AO, the signal attenuation will be significantly high with strong turbulence. 
Therefore, we assume a 1$\times$HV$_{5/7}$ Hufnagel-Valley turbulence profile \cite{sasiela2012electromagnetic}, and investigate how changing the FOV of the receiver changes performance.
Intuitively, since the algorithm seems highly resilient to noise, a potential strategy is to open the FOV in order to reduce channel attenuation beyond the 99$\%$ threshold.
The results are given in the downward progression in Figs.~\ref{fig:SUMMERnoAO} and \ref{fig:WINTERnoAO} where the FOV increases from 1$\times$ to 3$\times$ the diffraction limited FOV.
As one can see, increasing the FOV, which corresponds to a larger field stop in the focal plane, increases the sky hemisphere accessible by QTT, despite the increased probability of noise photons.

\subsection{Tracking and Higher Order AO}\label{sec:HEMISwAO}
\begin{figure}[]
	\includegraphics[width=0.9\columnwidth]{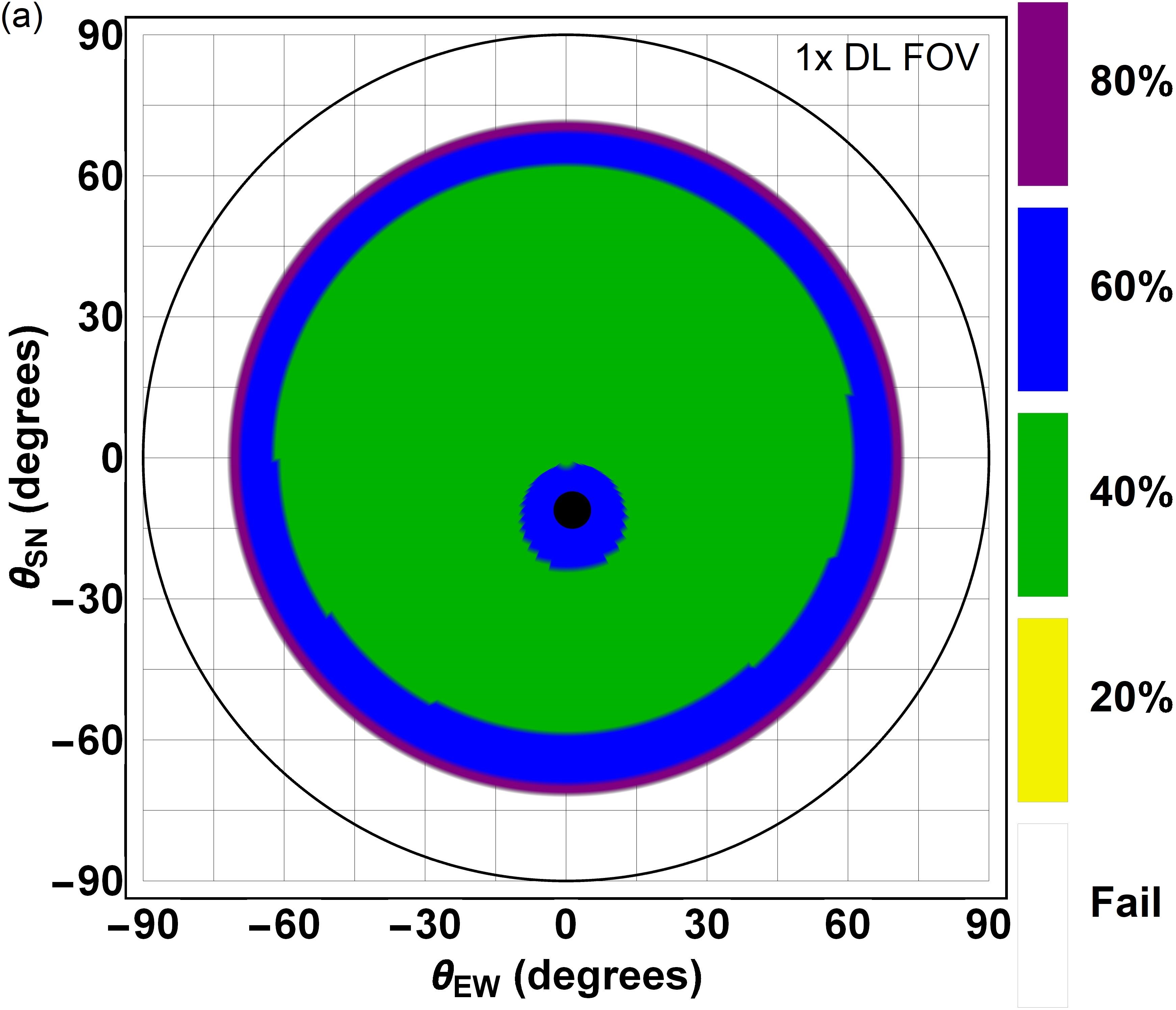}\\
	\vspace{0.0001cm} 
 	\includegraphics[width=0.9\columnwidth]{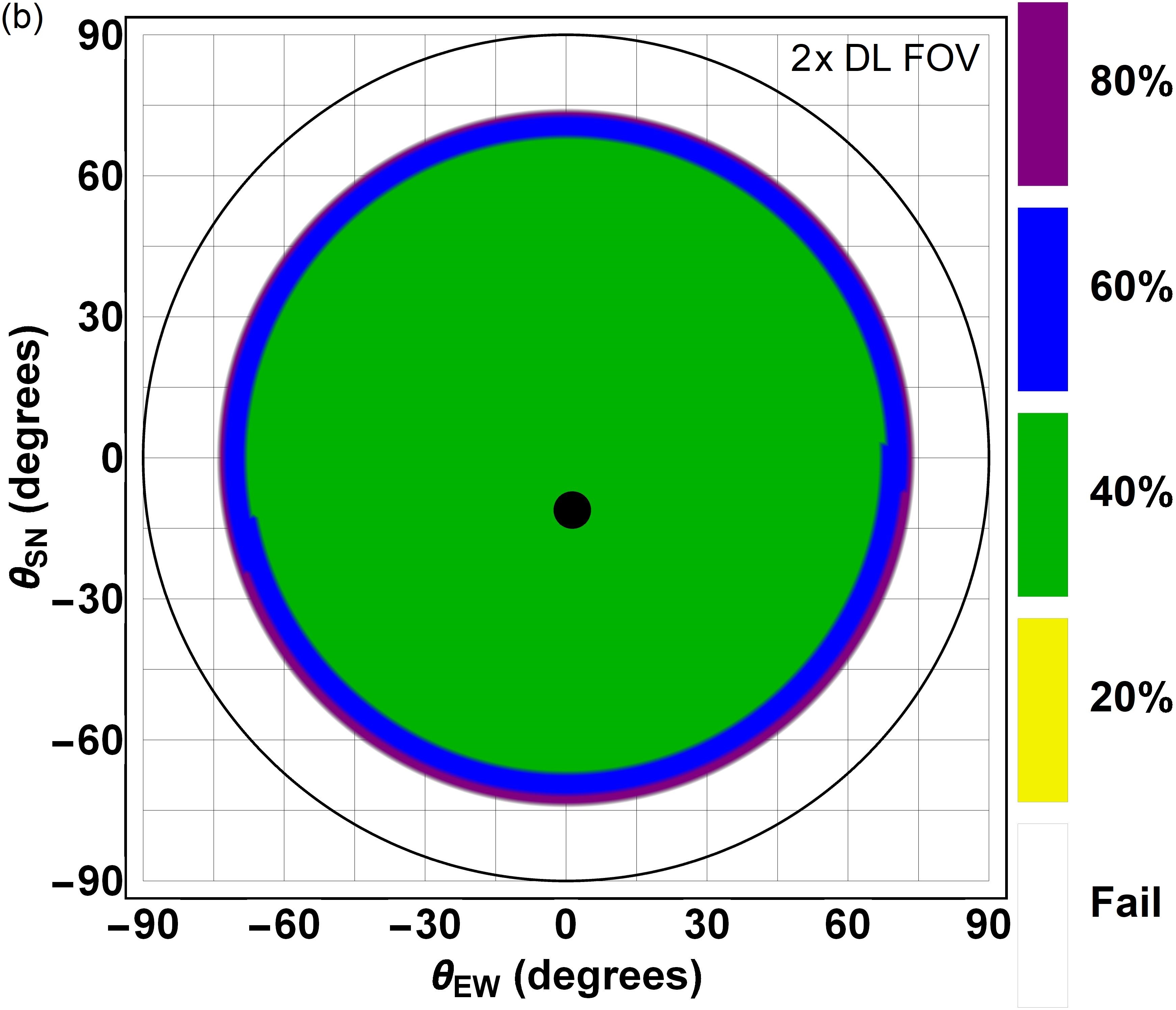}\\
 	\vspace{0.0001cm}
 	\includegraphics[width=0.9\columnwidth]{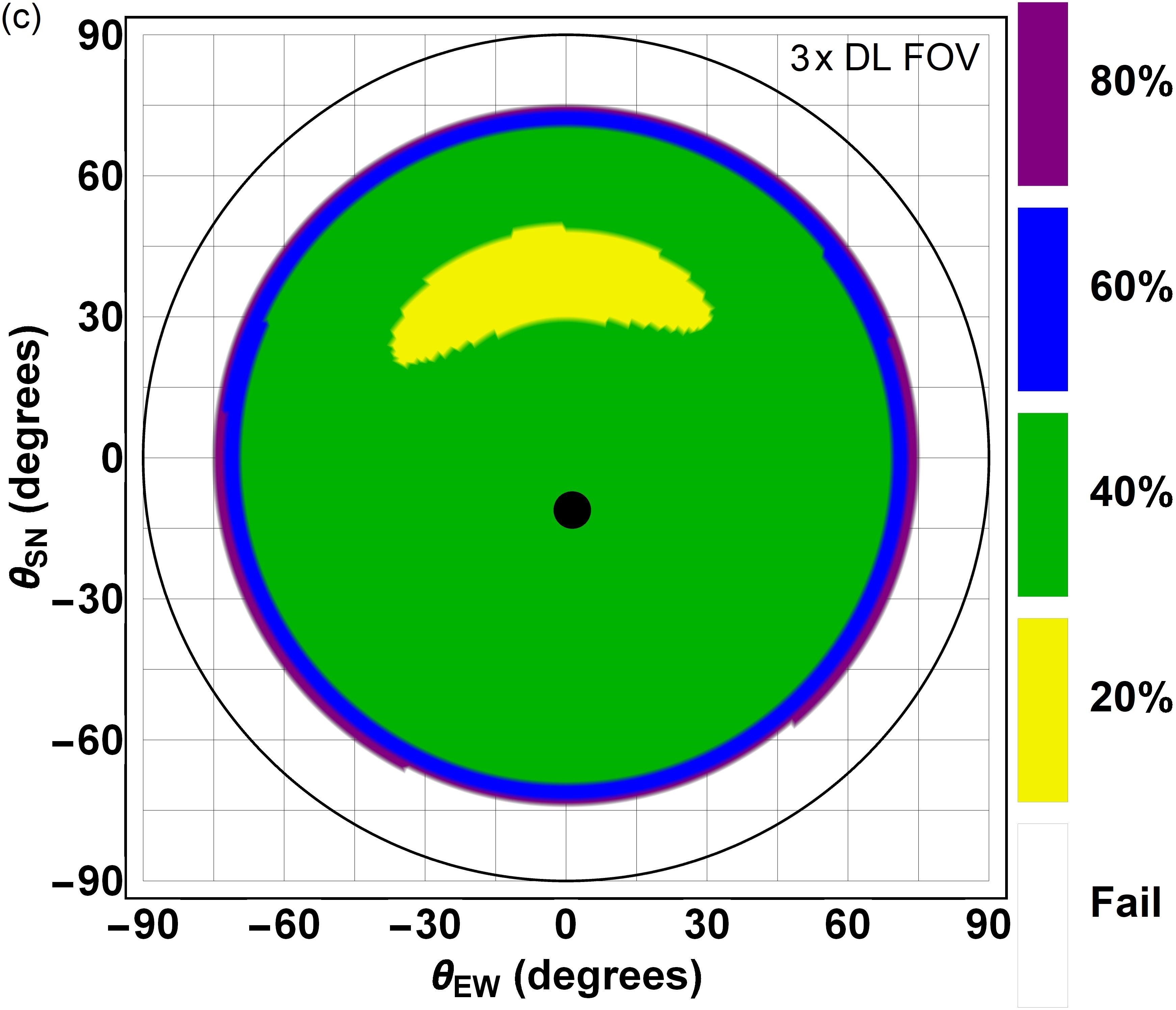}
	\caption{\label{fig:SUMMER}
Hemispherical plots of summer solstice with a 3$\times$HV$_{5/7}$ turbulence profile showing the regions where QTT succeeds with 99$\%$ probability for a receiver system with higher-order AO (see Sec.~\ref{sec:HEMISwAO} for more details). The legend gives the heralding efficiency of the biphoton source. In (a)--(c) we enlarge the FOV from 1$\times$ to 3$\times$ the diffraction limited FOV.
	}
\end{figure}

\begin{figure}[]
	\includegraphics[width=0.9\columnwidth]{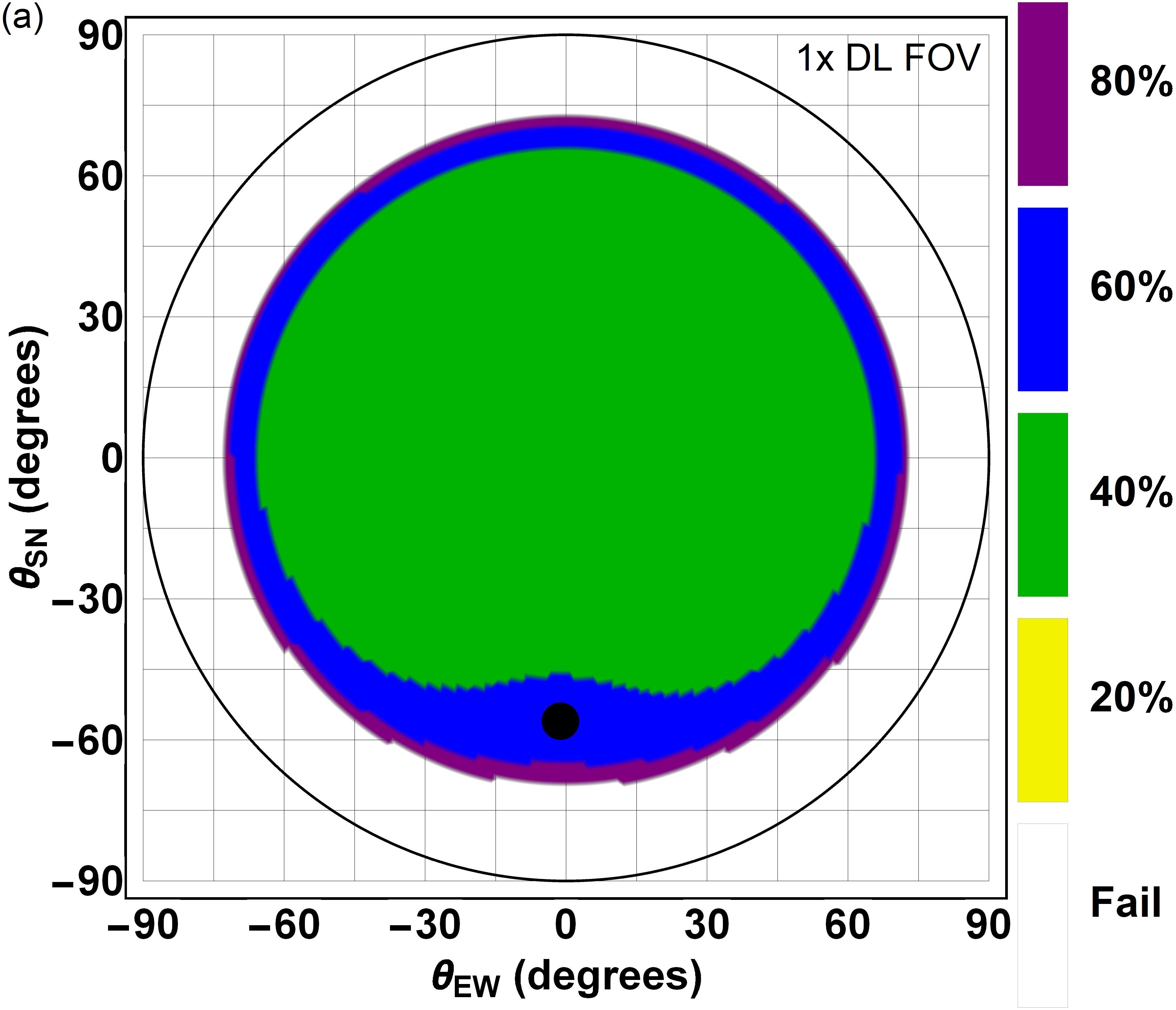}\\
	\vspace{0.0001cm} 
 	\includegraphics[width=0.9\columnwidth]{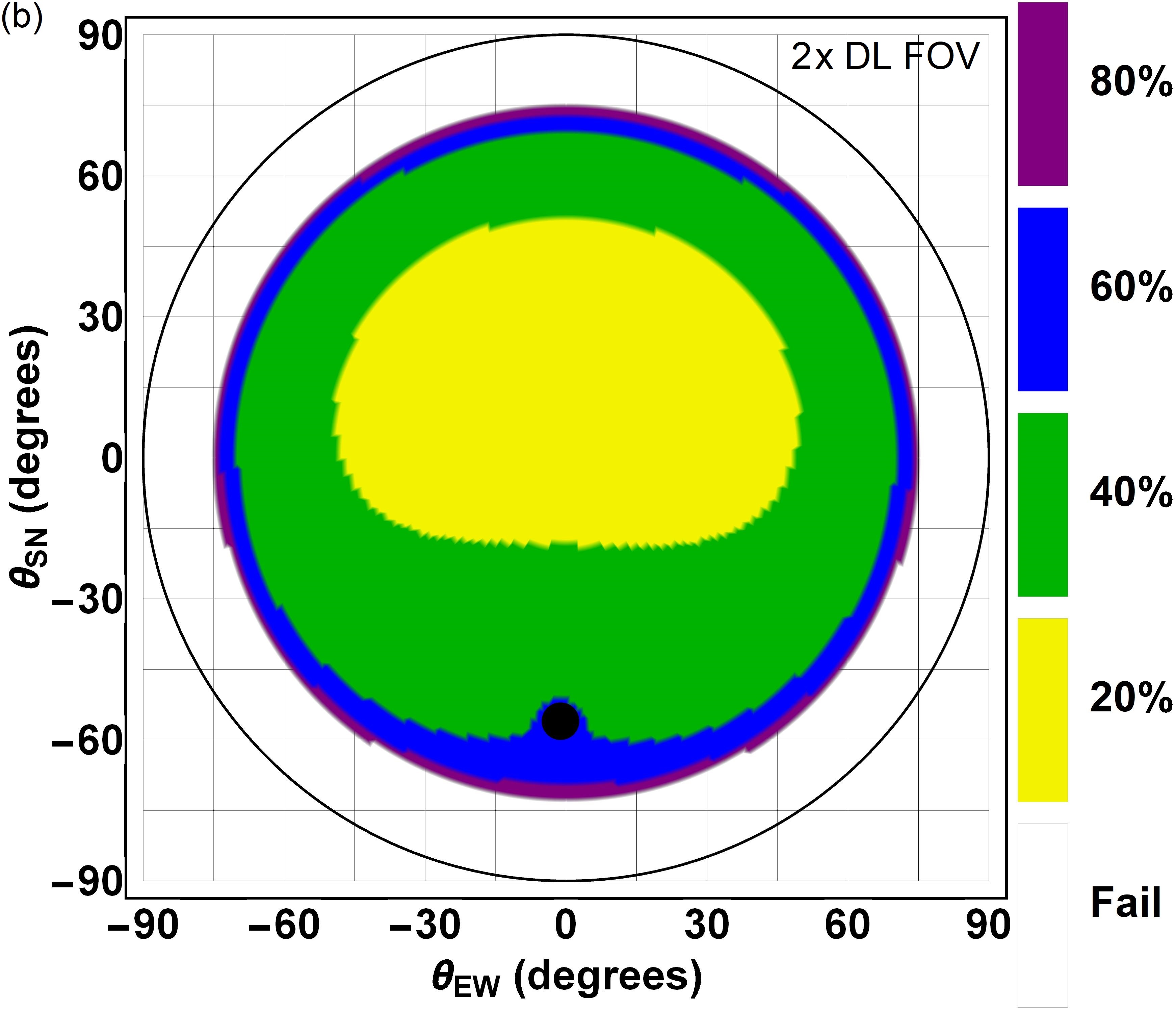}\\
 	\vspace{0.0001cm}
 	\includegraphics[width=0.9\columnwidth]{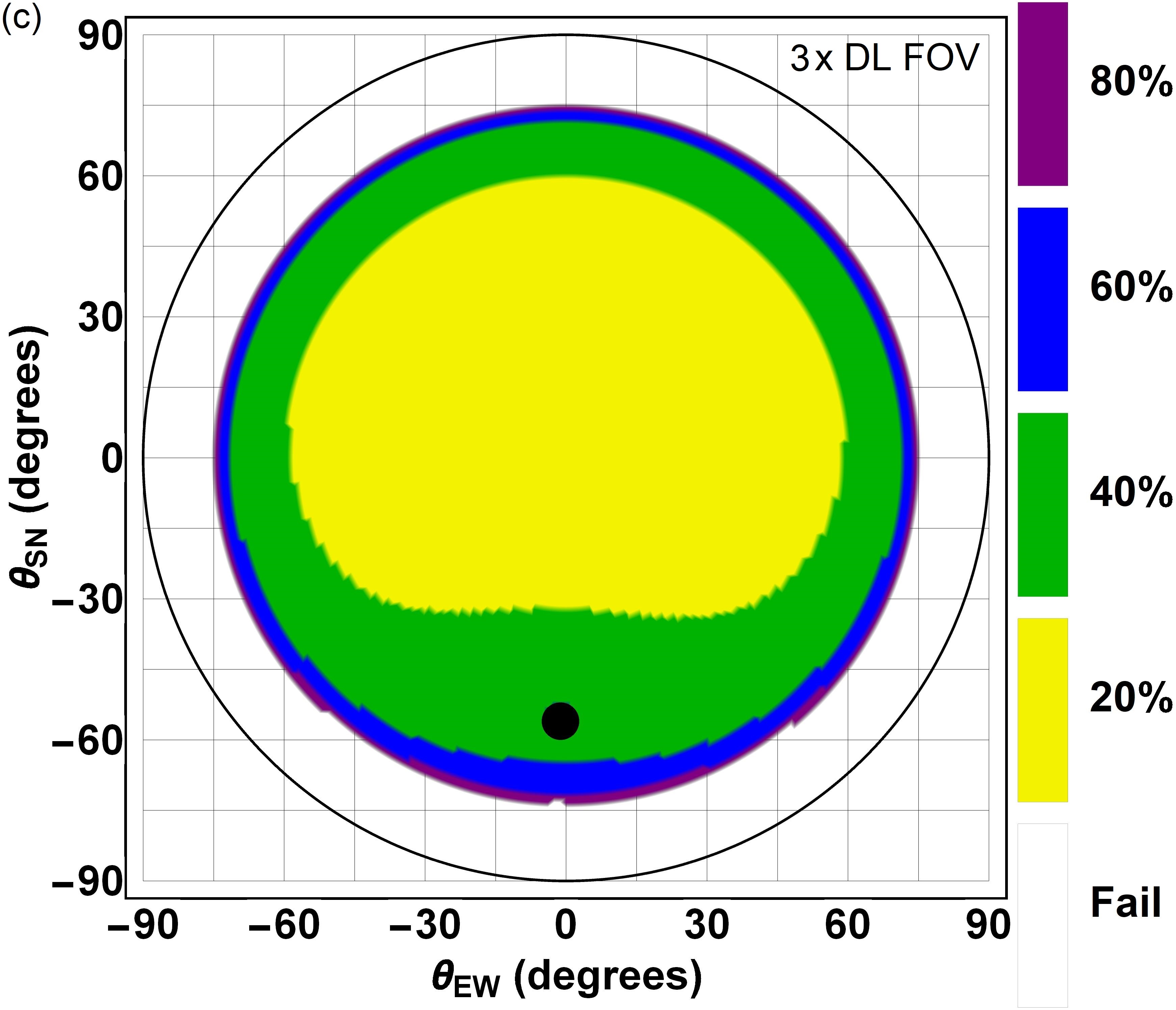}
	\caption{\label{fig:WINTER}
Hemispherical plots of winter solstice with a 3$\times$HV$_{5/7}$ turbulence profile showing the regions where QTT succeeds with 99$\%$ probability for a receiver system with higher-order AO (see Sec.~\ref{sec:HEMISwAO} for more details). The legend gives the heralding efficiency of the biphoton source. In (a)--(c) we enlarge the FOV from 1$\times$ to 3$\times$ the diffraction limited FOV.   
	}
\end{figure}

In the case of tracking and higher-order AO, one can estimate the residual error as a combination of the fitting error, aliasing error, and error due to the finite bandwidths of the tracking and higher-order AO systems \cite{hardy1998adaptive},
\begin{equation}\label{eq:RPEao}
\sigma_{\phi,\mathrm{CL}}^2 =1.3\times0.28 \Big( \dfrac{d_{\mathrm{sub}}}{r_0}  \Big)^{5/3}  + \Big(  \dfrac{\pi}{2} \dfrac{f_\mathrm{TG}}{f_{\mathrm{TC}}} \Big) ^{2} + \Big( \dfrac{f_\mathrm{G}}{f_c} \Big) ^{5/3},
\end{equation}
where $d_{\mathrm{sub}}$ is the subaperture spacing, that is matched to the DM actuator spacing, and the aliasing error is set to 30$\%$ of the fitting error.
With AO, a stronger and more realistic daytime turbulence strength can be considered.
In this case, we triple the turbulence strength by including a multiplicative factor on the turbulence profile, that is, we use a 3$\times$HV$_{5/7}$ Hufnagel-Valley profile \cite{sasiela2012electromagnetic}.
We perform the same investigation as the previous case and open the FOV in two steps.
Figures~\ref{fig:SUMMER} and \ref{fig:WINTER} give the results showing the considerable boost in performance that AO supports even in the case of 3$\times$ stronger turbulence.
It also shows that good performance can be achieved using sources available today, which can have heralding efficiencies in the range of $20$ to $40\%$, as long as AO is utilized.

In both cases, we have used conservative system parameters and restrictive constraints, e.g., 99$\%$ success probability, in order to demonstrate the utility of our methods. 
There is still quite a large trade space to be explored, and slight changes, for example larger aperture sizes, can make considerable changes to the results.
Nonetheless, this framework can be used to model many different link conditions and the simulation results can be applied to different link budgets.

\pagebreak
\clearpage
\newpage


\pagebreak
\clearpage
\newpage


\pagebreak
\clearpage
\newpage

A study exploring the extent to which opening the FOV is a suitable strategy should be included in any conceptual design.
Some noteworthy concerns are the effects of the system dead time as the rates get extremely high and the structure of the turbulence-broadened photon-probability distribution at the spatial filter. 

\section{Conclusion} \label{sec: Conclusion}
We investigate QTT as a candidate technique for the precise clock synchronization required to enable long-range daytime quantum networking and secure timing in GPS denied environments. 
The architecture is simple and the algorithm is robust to signal loss and the presence of noise photons.
We characterize the performance of QTT as a function of Bob's channel attenuation and the number of noise photons present in his channel.
Our results can subsequently be used to determine performance for specific design references given a link budget has been carefully derived.
We present the probability of successful time transfer and the standard error of the mean of the clock offset, which we show is the single-shot timing precision of the QTT algorithm.
We further show the standard error of the mean follows a $1/\sqrt{N_{\mathrm{T}}}$ trend, where $N_{\mathrm{T}}$ is the estimate of the number of ``true'' coincidences between Alice and Bob.
Setting a threshold probability of success to 99$\%$, our results show how many noise photons can be present and how many true coincidences to expect given a certain channel loss.
We also calculate the Allan deviation for different detection and clock systems, which conveys the stability and noise profile of the two clock system.
Finally, we interpret the results in the context of long fiber channels and daytime space-to-Earth downlinks, thereby demonstrating a specific design reference for the relevance of our method for global-scale quantum networking and timing in GPS denied environments.

In this work we model the timing jitter associated with clocks and hardware.
Thus, the results are valid for scenarios in which other deleterious effects have been modeled and compensated for, resulting in residual $\Delta U$ that is negligible.
Future work should include estimates of the residual $\Delta U$ after compensating for relative motion, and jitter caused by atmospheric fluctuations in ground-space links.
Similarly, future work could include modeling dispersion, temperature, and other effects unique to fiber networks.
It would also be interesting to investigate the performance of QTT in different configurations and across multiple nodes in a large network architecture.

\begin{acknowledgments}
The authors acknowledge helpful discussions with Scott Newey, Boeing Corporation, and program management support from Capt. Jonathan Schiller and Valerie Knight, AFRL.

The views expressed are those of the author and do not necessarily reflect the official policy or position of the Department of the Air Force, the Department of Defense, or the U.S. government.
The appearance of external hyperlinks does not constitute endorsement by the U.S. Department of Defense (DoD) of the linked websites, or the information, products, or services contained therein. 
The DoD does not exercise any editorial, security, or other control over the information you may find at these locations.
\end{acknowledgments}

\bibliography{quantum_time_transfer_TEM_v28}

\appendix

\section{Analytical Model} \label{sec:Analytic Model}
In this section we derive a model which predicts the performance of QTT based on parameters inferable from the channel conditions and hardware.
Ultimately, we derive an expression for the probability of success as a function of the correlation signal peak height.
We consider $N_T$ photon pairs randomly and uniformly distributed in an acquisition time $T_{a}$.
The detection time of each photon is randomized by the jitter of Alice and Bob's detectors, which we model as a Gaussian random variable.
Consequently, the detection-time difference histogram is a Gaussian with mean equal to the arrival time difference and standard deviation given by Eq.~\ref{eq:syserror}.

\subsection{Clock Drift}\label{sec:AppendixClockStability}
Next, we include the effect of the clock drift $\Delta U$.
We do this by dividing the acquisition time $T_{\mathrm{a}}$ into infinitesimal time increments $dT$.
During each time increment the detection times of the photon pairs are Gaussian distributed random variables with the same amplitude and standard deviation, but with center position shifted by $T \, \Delta U$.
Thus, the coincidence signal can be modeled as 
\begin{equation}
\label{eq:IntTa}
S =\frac{R_T}{\sigma_{\tau}\sqrt{2\pi}}\int_0^{T_{a}} \exp \Big[{-\frac{1}{2}\Big(\frac{\tau-T\Delta U}{\sigma_{\tau}}\Big)^2}\Big] dT,
\end{equation}
where $R_T$ is the true coincidence rate.
Integrating we find
\begin{equation}
\label{eq:AnalyticCorrHist}
S = \frac{R_T}{2\Delta U}\Big[\mathrm{erf}\Big(\frac{\tau}{\sqrt{2}\,\sigma_{\tau}}\Big)-\mathrm{erf}\Big(\frac{\tau -T_{a}\Delta U}{\sqrt{2}\,\sigma_{\tau}}\Big)\Big].
\end{equation}
To make this expression fit the simulation, or experimental results, one must accommodate for the effect of binning by multiplying Eq.~\ref{eq:AnalyticCorrHist} by $T_{\mathrm{bin}}$.
As a check, taking the limit $\Delta U \rightarrow 0$, one appropriately finds
\begin{equation}
S(\Delta U \rightarrow 0) = \frac{R_T T_a}{ \sigma_{\tau} \sqrt{2 \pi}}\exp \Big[{-\dfrac{1}{2} \Big(\dfrac{t}{\sigma_{\tau}}\Big)^2}\Big],
\end{equation}
noting that the product $R_T T_a = N_T$ is the number of true coincidences.
The expected number of detected true coincidences is 
\begin{equation}
\label{eq:TrueCoin}
N_T = N_{\mathrm{pair}}\eta_{\mathrm{herald}}^2\eta_A\eta_B\eta_{\mathrm{dead}}^A\eta_{\mathrm{dead}}^B,
\end{equation}
where $\eta_{\mathrm{herald}}$ is the heralding efficiency of the source, $\eta_A$ and $\eta_B$ are Alice and Bob's channel efficiencies, and $\eta_{\mathrm{dead}}^A$ and $\eta_{\mathrm{dead}}^B$ are the efficiencies due to detector dead time.

Using Eq.~\ref{eq:AnalyticCorrHist} we find the location and value of the signal peak
\begin{equation}
\begin{split}
\label{eq:Spk}
\tau_{\mathrm{peak}} &= \frac{1}{2}T_{a}\Delta U\\
S_{\mathrm{peak}} &= \frac{N_T}{T_a \Delta U}\mathrm{erf}\Big(\frac{T_{a}\Delta U}{2\sqrt{2} \, \sigma_{\tau}}\Big).
\end{split}
\end{equation}
Again, we can take the limit $\Delta U \rightarrow 0$ and appropriately find
\begin{equation}
\label{eq:Spk max}
S_{\mathrm{peak}}^{(\mathrm{max})} \equiv S_{\mathrm{peak}}(\Delta U \rightarrow 0)  =  \dfrac{N_T}{ \sigma_{\tau} \sqrt{2 \pi}}.
\end{equation}
Thus, we see that the effect of $\Delta U$ is to shift the peak and to reduce its height, which is detrimental to the precision and success probability of the QTT algorithm. 
Setting a threshold on the value of $S_{\mathrm{peak}}$, perhaps based on the noise distribution, limits $\Delta U$ relative to the acquisition time and jitter of the system.
For example, if one requires that $S_{\mathrm{peak}}$ be 99$\%$ of $S_{\mathrm{peak}}^{(\mathrm{max})}$, then
\begin{equation}
\Delta U \lessapprox \dfrac{1}{2} \dfrac{\sigma_{\tau}}{T_a}.
\end{equation}
If we assume SPAD detectors, then $\Delta U$ should be less than $2\times10^{-10}$ to meet the 99$\%$ $S_{\mathrm{peak}}^{(\mathrm{max})}$ threshold.

\subsection{Probability of Success}
One way to derive the probability of success is to calculate the probability that a noise peak could be misidentified as the correlation peak.
To do this, consider the background counts detected at Alice and Bob during the acquisition time $T_{\mathrm{a}}$.
The distribution of peak heights due to accidental coincidences has mean $\mu_b$ equal the average number of accidental coincidences per time bin
\begin{equation}
\label{eq:mub}
\mu_b = N_b^A N_b^B T_{\mathrm{bin}},
\end{equation}
where $N_b^A$ and $N_b^B$ are the observed background counts at Alice and Bob.
We find that it is sufficient to approximate the noise peak distribution $f(s)$ by a Gaussian with mean $\mu_b$ and standard deviation $\sigma_b=\sqrt{\mu_b}$,
\begin{equation}
f(s) = \dfrac{1}{\sigma_b\sqrt{2 \pi}} \exp\Big[-\dfrac{1}{2} \Big( \dfrac{s-\mu_b}{\sigma_b} \Big)^2 \Big].
\end{equation}
The probability that the peak is made up of accidental coincidences instead of the correlation signal is related to the $n$'th order statistic of the distribution of accidental peak heights $f(s)$, 
\begin{equation}
\label{eq:orderStat}
f_n(s) = nF(s)^{n-1}f(s),
\end{equation}
where $F(s)$ is the cumulative distribution function (cdf),
\begin{equation}
F(s) = \dfrac{1}{2} \Big[ 1 + \mathrm{erf} \Big(  \dfrac{s-\mu_b}{\sqrt{2} \, \sigma_b}  \Big)  \Big],
\end{equation}
of the distribution of accidental peak heights $f(s)$.
The order parameter $n$ is related to the number of misidentification opportunities in the correlation histogram.
The probability of success $P_s$ is the probability that the largest peak due to accidental coincidences is less than the maximum peak of the correlation histogram $S_{\mathrm{peak}}$.
\begin{figure}[t!]
	\includegraphics[width=0.9\linewidth]{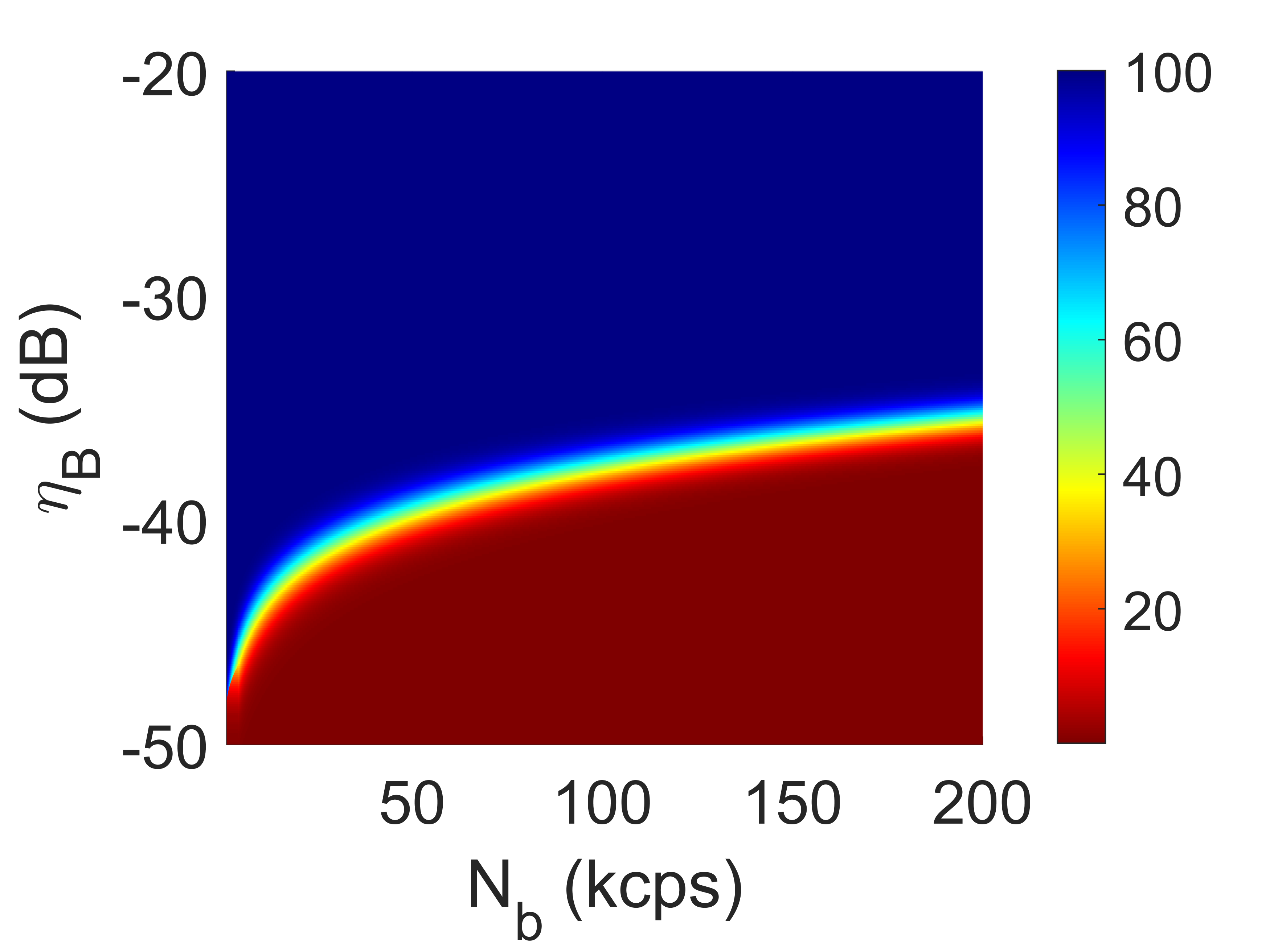}		
	\caption{Analytical probability of success using Eq.~\ref{eq:AnalyticProbability2} and 40$\%$ heralding efficiency. See Fig.~\ref{fig:prob_succ}(b) in the main text for the corresponding simulation result.}
	\label{fig:Analytical Prob S}
\end{figure}
Mathematically, this is the cdf of $n$'th order statistic $f_n(s)$ integrated to $S_{\mathrm{peak}}$, 
\begin{equation}
\label{eq:AnalyticProbability}
P_s = \int_0^{S_{\mathrm{peak}}} f_n(s)ds.
\end{equation}
Evaluating Eq.~\ref{eq:AnalyticProbability} one finds 
\begin{equation}
\label{eq:AnalyticProbability2}
P_s = 2^{-n}\Big[ \mathrm{erfc}\Big( \dfrac{\mu_b - S_\mathrm{peak}}{\sqrt{2} \, \sigma_b } \Big)^n  -  \mathrm{erfc}\Big( \dfrac{\mu_b }{\sqrt{2} \, \sigma_b } \Big)^n \Big].
\end{equation}
Eq.~\ref{eq:AnalyticProbability2} can be used to explain the shape of the contours in Fig.~\ref{fig:prob_succ} in the following way.
Inverting Eq.~\ref{eq:AnalyticProbability2} we find the threshold value of $S_{\mathrm{peak}}$
\begin{equation}
\begin{split}
S_{\mathrm{peak}}^{\mathrm{(thresh)}} &= \mu_b - \sqrt{2 \, \mu_b} \, \mathrm{erfc}^{-1}\bigg\{ \Big[ 2^n \\
&\times \Big(P_s + 2^{-n} \mathrm{erfc(\sqrt{\mu_b/2})^n}\Big) \Big]^{1/n} \bigg\}.
\end{split}
\end{equation}
Substituting in Eqs.~ \ref{eq:TrueCoin},\ref{eq:Spk}, and \ref{eq:mub} and solving for $\eta_B$, one can establish a function $\eta_B=\eta_B (P_s,N_b^B)$ and predict the shape of the contours in Fig.~\ref{fig:prob_succ}.
Keeping the first two terms in a Taylor series expansion one finds an expression of the form
\begin{equation}
\label{eq:ThrAtt}
\begin{split}
\eta_B = c_2 \sqrt{N_b^B} + c_1 N_b^B,
\end{split}
\end{equation}
where
\begin{equation}
\begin{split}
c_2 &= -\dfrac{\sqrt{2 N_b^A \, T_{\mathrm{bin}}} \, (T_a  \Delta U)  \, \mathrm{erfc}^{-1}(\mathcal{P}^{1/n})}{\mathcal{N}},\\
c_1 &= \dfrac{(N_b^A T_{\mathrm{bin}}) \, (T_a \Delta U) \Big(  \mathcal{P} - \exp\Big[ \mathrm{erfc}^{-1}(\mathcal{P}^{1/n}) \Big]^2 \mathcal{P}^{1/n} \Big)}{\mathcal{P} \,\mathcal{N}},\\
\mathcal{N} &= N_{\mathrm{pair}} \eta_A \eta_A^{\mathrm{dead}} \eta_B^{\mathrm{dead}} \eta_{\mathrm{herald}}^2 \mathrm{erf}\Big(  \dfrac{T_a \Delta U}{2\sqrt2 \sigma_{\tau}} \Big),\\
\mathcal{P} &= 1+2^n P_s.
\end{split}
\end{equation}

This analytical approach correctly predicts the overall shape observed from the Monte Carlo simulation, but it generally overestimates the performance of QTT.
In other words, the threshold attenuation predicted by Eq.~\ref {eq:ThrAtt} is shifted downward compared to our simulation results.
For example, in Fig.~\ref{fig:Analytical Prob S} we plot the analytical probability of success with 40$\%$ heralding efficiency.
We choose the order parameter $n=14$ and see that the analytical approach overestimates the performance by $\sim$2~dB.
The discrepancy suggests the model may be incomplete.
For example, the systematic steps of the proposed algorithm, such as the peak-finding process, are not accommodated for in the this model.
Fundamentally, the signal peak height is simply \textit{not} an unbiased estimator of the population peak height, which makes modeling the algorithm inherently problematic. 
Nonetheless, we find this progression useful because it justifies the fit used to trace out the 99$\%$ threshold curves in Fig.~\ref{fig:prob_succ} in the main text and can be used to estimate performance.

\end{document}